\documentclass[aps,floatfix,twocolumn,prb,superscriptaddress]{revtex4-2}
\usepackage{xcolor} 
\usepackage{graphicx}
\usepackage[hidelinks]{hyperref}
\usepackage{amssymb}
\usepackage[utf8]{inputenc}
\usepackage{amsmath}
\usepackage{braket}
\usepackage{adjustbox, makecell}
\newcommand{\abs}[1]{\left|#1 \right|}
\newcommand{\mat}[1]{\mathrm{#1}}
\newcommand{\op}[1]{\mathrm{#1}}

\renewcommand{\vec}[1]{\boldsymbol{#1}}
\newcommand{\iu}[0]{\mathrm{i}}

\definecolor{vastkust}{RGB}{0, 48, 80} 
\hypersetup{
  colorlinks,
  citecolor=vastkust,
  linkcolor=vastkust,
  urlcolor=vastkust}

\begin{document}
\title{Axial phono-magnetic effects}
\author{Natalia\ Shabala}
\email{natalia.shabala@chalmers.se}
\author{Finja\ Tietjen}
\email{finja.tietjen@chalmers.se}
\author{R.~Matthias\ Geilhufe}
\email{matthias.geilhufe@chalmers.se}
\affiliation{Department of Physics, Chalmers University of Technology, 412 96 G\"{o}teborg, Sweden}
\date{\today}
\begin{abstract}
Axial or circularly polarized phonons are collective lattice vibrations with angular momentum. Over the past decade they have emerged as a promising mechanism for the manipulation of magnetism, in parallel to well established optical protocols. In particular, coherent axial phonons were shown to induce magnetization in materials without spin-ordering, making them a viable tool for ultrafast magnetic switching. The experimental evidence suggests that the size of this magnetization is significant, opening a new research area on the phono-magnetic effect. Remarkably, the coupling of axial phonons to magnetism has been observed a broad class of materials, pointing to a universal nature of the underlying mechanisms. In this review article, we present the recent progress in the field.
We give an introduction to the phenomenological perspective and an overview of the experimental evidence for the magnetization emerging from axial phonons, which includes discussing the observations of phonon Zeeman effect, the magneto-optical Kerr effect and the proximity-induced magnetization switching.
We present recently proposed microscopic theories for the phono-magnetic effects, based on perturbation theory, adiabatic motion and Floquet theory as well as the emergence of the phonon magnetic moment due to artificial gauge fields or inertial effects. This summary allows us to see correspondences between the seemingly different theoretical approaches, facilitating a more complete perspective of the effect.
\end{abstract}
\maketitle

\tableofcontents

\section{Introduction}
Magnetism and its manipulation have played a crucial role in nanotechnology, particularly in areas such as data storage, information processing, and spintronics. 
In this context, the control of magnetic order through light has been a major focus of research, ranging from the Faraday and inverse Faraday effects to more recent studies on ultrafast demagnetization experiments and nonlinear magnonics~\cite{pershan1966theoretical, Beaurepaire1996, Kirilyuk2010, Zheng2023, schonfeld2023}. 
Initially, phonons, which are collective lattice excitations, were considered to play a minor role and were primarily discussed in the context of heat or the renormalization of magnetic excitations through phonon-magnon coupling~\cite{Kim2018, George1970, Wang2023}.

However, in recent years, the perception of the role of phonons in magnetism and magnetic manipulation experiments has radically changed, primarily due to two guiding principles. First, optical phonons have the same symmetry as light. Therefore, almost all types of light-matter interaction have a phononic counterpart~\cite{Juraschek2020phono}. Second, phonons were shown to carry angular momentum~\cite{zhang2014angular,Coh2023,zhang2025chirality,Zhang2025} together with an effective charge~\cite{Chen1968}. As a result, phonons carry a magnetic moment that interacts with magnetic fields and the spin of the electron. 

Initially, the phonon magnetic moment was approximated by the classical motion of an ionic charge and estimated to be on the order of the nuclear magneton $\mu_N$, which differs from the magnetic moment of an electron (Bohr magneton $\mu_B$) by the electron-to-proton mass ratio, $\mu_N \approx 1/1836~\mu_B$~\cite{rebane1983faraday,juraschek2017dynamical,juraschek2019orbital,geilhufe2021dynamically}. However, various experiments have undoubtedly shown that the measured phonon magnetic moment is about 3-4 orders of magnitude larger, and in fact closer to the Bohr magneton~\cite{Schaack1975,Schaack1976,Schaack1977,cheng2020large,baydin2022magnetic,hernandez2023observation,luo2023,Basini2024,Davies2024}. Furthermore, this phenomenon seems to be universal and observed for various material classes, such as Dirac semimetals, insulators, topological materials, as well as paraelectrics.

Inspired by the compelling experimental findings, a bulk of theoretical work has been developed, explaining the emergence of the large phonon magnetic moment predominantly by coupling phonons and electrons. An angular momentum transfer from phonons to electrons allows to exploit the electron-to-ion mass ratio and significantly increases the gyromagnetic ratio. Typically, theoretical work focuses on a specific method and approximation and is often shown in the example of a specific material. It is the aim of the present review to summarize the ongoing theoretical work and put it on a more general footing. This allows us to see the equivalence between seemingly distinct formulations.

We note that overlapping terminology has been introduced in the literature, relating to individual subject areas. For example, inspired by the optical inverse Faraday effect, the term phonon inverse Faraday effect has been introduced to describe the magnetization of a circularly polarized optical phonon~\cite{rebane1983faraday}. Similarly, an optical phonon can be regarded as a fluctuating polarization. This fluctuating polarization $\vec{P}$ induces a magnetization in the form of $\vec{M}\sim \vec{P}\times\dot{\vec{P}}$, giving rise to the term dynamical multiferroicity~\cite{juraschek2017dynamical}, drawing the connection to multiferroics, i.e., materials with coexisting polarization and magnetization. 
Additionally, phono-magnetic effects are connected to magneto-mechanical phenomena like the Barnett effect~\cite{Barnett1915}, describing magnetization in a spinning uncharged body. Recently, this concept has been applied to the atomic scale, where circularly polarized phonons create magnetization termed the ultrafast Barnett effect~\cite{Davies2024,Basini2024}. Conversely, the Einstein--de Haas effect~\cite{EinsteindeHaas1915} has been used to describe ultrafast angular momentum transfer from electronic spin to phonons~\cite{Tauchert2022}. However, we note that all these terminologies are closely related or even equivalent and should depend on a unified theoretical principle.  

Finally, there have been differing notations on the phonons themselves, using circularly polarized phonons, axial phonons, and chiral phonons interchangeably. This matter was recently addressed with the following definitions~\cite{juraschek2025chiral}: chiral phonons~\cite{Zhang2015,zhu2018observation,ishito2023truly} are phonons that break improper rotation symmetry, whereas axial phonons are phonons that carry real~\cite{zhang2014angular,juraschek2017dynamical,geilhufe2021dynamically} and/or pseudo angular momentum~\cite{Zhang2015,zhang2022chiral,zhang2025chirality}. Circularly polarized phonons carry angular momentum and are regarded as axial. In contrast, chirality and axiality are not equivalent, giving rise to geometric chiral phonons~\cite{Romao2024,Zhang2024under,Fava2025,Zeng2025} (chiral but not axial), axial chiral phonons, and axial achiral phonons. Here, we focus on emergent magnetic effects due to phonon angular momentum, i.e., axial phonons.

To shed light on the axial phono-magnetic effect, we use the following structure for our review. 
We begin by introducing phonons and phonon angular momentum. We continue by motivating the emergence of phono-magnetism using a phenomenological Landau theory as well as the concept of dynamical multiferroicity, where magnetization emerges from time-varying polarization. Afterwards, we give a concise summary of experimental observations of the phonon magnetic moment, through techniques such as the phonon Zeeman effect, magneto-optical Kerr effect (MOKE), and proximity-induced magnetization switching. Afterwards, we delve into microscopic theories that explain these effects, categorizing them into perturbative, adiabatic, and Floquet approaches, and emphasizing the critical role of electron-phonon interactions. The review further examines inertial effects, such as spin-rotation coupling, and their contribution to magnon-phonon hybridization. We conclude by summarizing the relationship found between seemingly different theoretical frameworks and provide an outlook on potential technological applications and future research directions.

We note that the field of phonomagnetism has been growing rapidly in the process of finalizing this review. Hence, we were not able to cover closely related research areas, where axial phonons are described using ab initio methods, such as time-dependent density functional theory~\cite{Mrudul2025}, spin-lattice coupling~\cite{weissenhofer2024truly, Miranda2025,hellsvik2019general}, velocity-force coupling~\cite{bonini2023frequency}, or theoretical methodology connected to exact factorization~\cite{requist2019exact,abedi2010exact}. Furthermore, similar methodology to axial phonons has been developed in adjacent fields, e.g. discussing the emergence of circular modes in metamaterials~\cite{Benzoni2021,Marijanovi2022}.

\section{Phenomenology}

\subsection{Axial phonons and phonon angular momentum}\label{section:dynamical_multiferroicity}
Crystalline materials are composed of a multitude of ions arranged in a periodic lattice. Each atom can be assigned a position $\vec{R}_{l\alpha}$, with $l$ running over all unit cells and $\alpha$ running over the sites within each unit cell. At finite temperature, the ions fluctuate around their equilibrium positions $\vec{R}^{(0)}_{l\alpha}$ with a small displacement $\vec{\tau}_{l\alpha}$,
\begin{equation}
   \vec{R}_{l\alpha} = \vec{R}^{(0)}_{l\alpha} + \vec{\tau}_{l\alpha}.
   \label{eq:disp}
\end{equation}
Phonons are collective lattice excitations and therefore described in terms of the $\vec{\tau}_{l\alpha}$. Due to the lattice periodicity, it is convenient to transform into the Fourier space, $\vec{\tau}_{l\alpha} \rightarrow \vec{\tau}_{\vec{k}\alpha}$. Furtherore it is common to incorporate the ionic mass $M_\alpha$ and define,
\begin{equation}
    \vec{u}_{\vec{k}\alpha} = \sqrt{M_\alpha} \vec{\tau}_{\vec{k}\alpha}.
\end{equation}
The classical equation of motion for $\vec{u}_{\vec{k}\alpha}$ follows from Newton's equation of motion and, in the harmonic approximation, is given by a set of coupled harmonic oscillators, 
\begin{equation}
      \ddot{u}_{\vec{k}\alpha i}  = - \sum_{\alpha' j} D_{\alpha i, \alpha' j}(\vec{k}) u_{\vec{k} \alpha' j}.
\end{equation}
Here, $D_{\nu\alpha, \nu'\beta}(\vec{k})$ is the dynamical matrix. The Cartesian coordinates are indexed by $i$ and $j$. As the site indices $\alpha,\beta$ run from $1,\dots,N$, with $N$ being the number of atoms in the unit cell, the dynamical matrix has dimension $3N\times 3N$. As a result, one obtains $3N$ eigenvectors, i.e., phonon modes $u_{\vec{k}\nu}$, $\nu = 1,\dots,3N$, for each wave vector $\vec{k}$ in the Brillouin zone. The eigenvalues of the dynamical matrix are the squares of the phonon frequencies $\omega_{\vec{k}\nu}^2$.

Axial phonons are phonons carrying phonon angular momentum. Over the past years, two types of phonon angular momentum have been introduced: real phonon angular momentum and pseudo-phonon angular momentum. Real phonon angular momentum $\vec{J}^{\text{ph}}$ follows the classical definition of angular momentum, as proposed by Zhang and Niu \cite{zhang2014angular}. While lattice angular momentum arises from the rotation of the crystal itself, phonon angular momentum has its origins in lattice vibrations. In real space, it is formulated as 
\begin{equation}
    \vec{J}^{\text{ph}} = \sum_{l\alpha} M_\alpha\, \vec{\tau}_{l \alpha} \times \dot{\vec{\tau}}_{l \alpha} = \sum_{l\alpha} \vec{u}_{l \alpha} \times \dot{\vec{u}}_{l \alpha}\, .
\end{equation}
Such phonon angular momentum can be introduced, e.g., by coherent laser excitation~\cite{Basini2024,Davies2024}, phonon-phonon scattering~\cite{Minakova2025}, ultrafast demagnetization~\cite{Tauchert2022}, or thermal gradients \cite{Zhang2025}. Furthermore, it should be inherently present in magnetic materials with broken time-reversal symmetry.

So far, we have introduced the concept of phonons through classical ionic displacements. However, similar to the photon, which represents the quantized electromagnetic field, the proper definition of a phonon is the quantized lattice vibration, based on the quantized displacements
\begin{equation}
    \tau_{l \alpha i} = \sqrt{\frac{M_0}{M_\alpha N_l}} \sum_{\vec{q}\nu} e^{i\vec{q}\cdot \vec{R_l}}e_{\nu, \alpha i}(\vec{q}) l_{\vec{q}\nu}(\hat{a}_{\vec{q}\nu}+\hat{a}^{\dagger}_{-\vec{q}\nu}).
    \label{eq:quantizeddisp}
\end{equation}

Here, $M_0$ is the reference mass, $N_l$ the number of cells, $l_{\vec{q}\nu} = \sqrt{\hbar/(2M_0 \omega_{\vec{q}\nu})}$ the zero mode displacement, and $e_{\nu, \alpha i}$ the normalized $\nu^{\text{th}}$ eigenvector of the dynamical matrix. As a result, the quantized phonon angular momentum can be written as follows~\cite{zhang2014angular},

\begin{equation}
    J_{ph} = \sum_{\mathbf{k} \nu}  L_{\mathbf{k}\nu}\left[n(\omega_{\mathbf{k},\nu})+\frac{1}{2}\right].
    \label{eq:realangular}
\end{equation}

Here, $n(\omega_{\mathbf{k},\nu})$ is the occupation number of the mode $\omega_{\mathbf{k},\nu}$, given by the Bose-Einstein distribution in thermal equilibrium. Similar to the finite ground energy of the harmonic oscillator, the quantized phonon angular momentum is finite for $n(\omega_{\mathbf{k},\nu})=0$ and given by the mode angular momentum $L_{\mathbf{k}\nu}$.

Phonons can also carry pseudoangular momentum, a distinct physical quantity from conventional angular momentum. For a phonon Bloch wave function $u_{\nu\mathbf{k}}e^{i \mathbf{k}\cdot\mathbf{R}_{l\alpha}-i\omega t}$, which is invariant under $n$-fold rotation, the pseudoangular momentum $l_{\text{ph}}$ is defined as

\begin{equation}\label{eq:phonon_PAM}
    \mathcal{C}_n u_{\nu\mathbf{k}}e^{i \mathbf{k}\cdot\mathbf{R}_{l\alpha}} = e^{-i\frac{2\pi}{n}l_{\text{ph}}} u_{\nu\mathbf{k}}e^{i \mathbf{k}\cdot\mathbf{R}_{l\alpha}}.
\end{equation}

Here, $\mathcal{C}_n$ is the $n$-fold rotation operator \cite{Zhang2015, zhang2022chiral, zhang2025chirality}.

The pseudoangular momentum depends on the invariance of the phonon wave function under $n$-fold rotation and is determined by the rotational symmetry of the crystal. At high-symmetry points in reciprocal space, where $\mathbf{k}$ is invariant under the rotation operator $C_n$, $l_{\text{ph}}$ can take values of $0,\dots,n-1$ in symmorphic crystals and fractional values in nonsymmorphic crystals \cite{zhang2022chiral}. In 2D and 3D crystals, only $n=\{2,3,4,6\}$-fold rotations are allowed.

Similarly to spin and orbital angular momentum of light, it is possible to define spin $l_{s}$ and orbital $l_{o}$ pseudoangular momentum of phonons. This is a consequence of the rotation operator acting separately on the intracell part (spin) and the intercell part (orbital) of the phonon wave function. The total pseudoangular momentum is a sum of both contributions \cite{Zhang2015}

\begin{equation}
   l_{\text{ph}}=l_{s}+l_{o}.
\end{equation}

While the spin angular momentum is connected to real mechanical angular momentum, the orbital angular momentum is connected to scattering selection rules \cite{zhang2025chirality}. The concept of pseudoangular momentum has been central in the early definition of chiral phonons \cite{Zhang2015} and their experimental observations in 2D \cite{zhu2018observation} and 3D \cite{ishito2023truly}.
It has also been shown that phonon modes possessing pseudoangular momentum can provide a way to probe multipolar magnetic ordering, i.e., octupolar or quadrupolar ordering \cite{Sutcliffe2025}.

\subsection{Dynamical multiferroicity}

\begin{figure}
    \centering
    \includegraphics[width=0.66\linewidth]{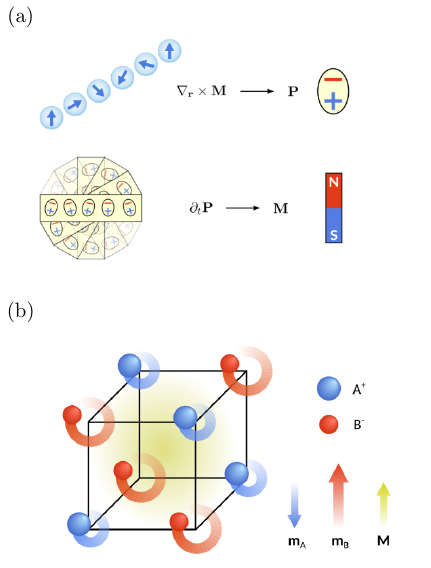}
    \caption{Schematic representation of dynamical multiferroicity. a) Two reciprocal processes of generating multiferroicity: through spatially varying magnetization or temporally varying polarization. b) Magnetization induced by collective circular motion of ions: two types of ions create local magnetic moments of different magnitude which leads to non-zero net magnetization.
    \textit{Reproduced with permission from~\cite{juraschek2017dynamical}; Copyright 2017 American Physical Society.}
    }
    \label{fig:dynamical_multiferroicity}
\end{figure}

Multiferroics are materials with coexisting magnetization and polarization~\cite{Khomskii2006,Tokura2014,Fiebig2016,Spaldin2019,Bossini2023}. The concept of multiferroics can be extended into the time domain, giving rise to dynamical multiferroicity~\cite{juraschek2017dynamical}. Here, magnetization arises through a temporarily varying polarization $\vec{P}$ \cite{juraschek2017dynamical}, with the magnetic moment expressed by 
\begin{equation}\label{multiferroicity}
    \vec{\mu} \sim \vec{P} \times \partial_t \vec{P},
\end{equation}
Dynamical multiferroicity can be viewed as a process reciprocal to the Dzyaloshinskii-Moriya interaction where a spatially varying magnetization induces electric polarization, as schematically depicted in Fig. \ref{fig:dynamical_multiferroicity}. Furthermore, an ionic displacement introduces a local charge in the crystal, giving rise to the polarization
\begin{equation}
    P_i = \frac{1}{\Omega_0}\sum_{\alpha j} Z_{\alpha,ij} \,u_{\vec{0}\alpha j} = \frac{1}{\Omega_0}\sum_\nu Z_{\nu i} \,u_{\vec{0}\nu},\label{eq:BornZ}
\end{equation}
with $\Omega_0$ denoting the unit cell volume. $Z_{\alpha,ij} = \Omega_0 \frac{\mathrm{d}P_i}{\mathrm{d}u_{\vec{0}\alpha j}}$ refers to the Born effective charge and $Z_{\nu j}$~\cite{Ghosez1998,Gonze1997} the mode effective charge.

Since optical phonons are polar, lattice vibrations can result in the magnetization of an otherwise non-magnetic material. Equation \eqref{multiferroicity} can be written in terms of normal coordinates $\vec{u} = \left(u_{\vec{0}1},u_{\vec{0}2}\right)$ of (degenerate) optical $\Gamma$-point phonons ($\vec{q}=\vec{0}$):
\begin{equation}
    \mu = \gamma \, \left(u_{\vec{0}1}\dot{u}_{\vec{0}2}-u_{\vec{0}2}\dot{u}_{\vec{0}1} \right) = \gamma J^{\text{ph}}.
    \label{eq:phonomag}
\end{equation}
Thus, as in classical electrodynamics, the magnetization is a product of the phonon angular momentum $\vec{J}^{\text{ph}} = \vec{u} \times \vec{\dot{u}}$ with the gyromagnetic ratio $\gamma$~\cite{juraschek2017dynamical}. 
However, it is worth noting that phonon magnetic does not always follow the proportionality described by equation \eqref{eq:phonomag}. 
It has been shown that phonon magnetic moment can arise from the phonon pseudoangular momentum alone, from an almost vanishing phonon angular momentum in combination with a diverging gyromagnetic ratio, or from angular momentum that is not parallel to the resulting phonon magnetic moment \cite{chaudhary2025anomalous}.
These phenomena represent anomalous phonon magnetic moments. In this review, we restrict our discussion to the conventional phonon magnetic moment as defined by equation \eqref{eq:phonomag}.

Phonon angular momentum can be induced by subjecting the system to excitation by a circularly polarized THz laser field. Simulations of the corresponding phonon dynamics are typically done using classical equations of motion. Using the complex phonon mode $u = u_{\vec{0}1} + i u_{\vec{0}2}$ they can be written in the compact form 
\begin{equation}
    \ddot{u} + \eta \dot{u} + \omega_0^2 u = \beta\, Z E(t),
    \label{eq:phonoeqm}
\end{equation}
with $\eta$ a phenomenological damping parameter, $\omega_0$ the mode frequency, $Z$ the mode effective charge tensor, $E(t)$ the time-dependent electric field acting on the mode, and $\beta$ a phenomenological screening parameter. Hence, for an electric field of the form $E(t) = f(t) e^{i \omega t}$ and a real-valued amplitude $f(t)$, both phonon modes are oscillating with a relative phase of $\pi/2$, resulting in a non-zero angular momentum.

Using this approach, and realistic laser field strengths of up to 1~MV/cm$^{-1}$~\cite{Saln2019}, the excitation of coherent phonon modes and the expected phonon magnetic moment were calculated for various materials using the paradigm of dynamical multiferroicity~\cite{juraschek2017dynamical,juraschek2019orbital,geilhufe2021dynamically} (see Figure~\ref{fig:dynamical_multiferroicity}). Accounting for typical ionic charges of a few elementary charges, displacements in the fractions of an \AA ngström, together with typical ionic masses, restricts the size of the phonon magnetic moment to the order of the nuclear magneton, i.e., $\approx 5.4 \times 10^{-4}~\mu_B$. 
 
\subsection{Landau theory}\label{subsec:Landau_theory}
On the mesoscopic scale, the phonon-induced magnetism can be derived using Landau theory~\cite{shabala2024phonon}. The formalism is equivalent to the optical analogue - the inverse Faraday effect~\cite{pershan1966theoretical} - giving rise to the notion of the phonon inverse Faraday effect.

Let us consider a system with two optical $\Gamma$-point phonon modes with mode amplitudes $u_\mu$ and $u_\nu$ and phonon frequency $\omega$. Then, the total ionic displacement can be written as:
\begin{align}
    \vec{u}(t) &= \left(u_\mu(t) \vec{\hat{e}}_\mu + u_\nu(t) \vec{\hat{e}}_\nu\right)e^{i \omega t} \\
    &= \left(\frac{1}{\sqrt{2}}u_R(\vec{\hat{e}}_\mu + i\vec{\hat{e}}_\nu) + \frac{1}{\sqrt{2}} u_L(\vec{\hat{e}}_\mu - i\vec{\hat{e}}_\nu)\right)e^{i\omega t},
    \label{circularphonon}
\end{align}
where the second line corresponds to a basis transformation into right and left polarized axial phonon modes $u_R = \left(u_\mu - i u_\nu\right)/\sqrt{2}$ and $u_L = \left(u_\mu + i u_\nu\right)/\sqrt{2}$. 

The magnetization can be obtained by taking the derivative of the free energy function $F$ with respect to the magnetic field $\vec{B}$,
while the free energy function can be constructed from symmetry properties. In the simplest case, we assume a non-magnetic and centrosymmetric crystal, which consequently is even under time reversal and inversion symmetries. As phonon modes are collective displacements, $\vec{u}(t)$ follows the transformation behavior of a conventional vector, being odd under inversion and even under time reversal symmetry. In contrast, magnetization $\vec{M}$ and the magnetic field $\vec{B}$ are pseudo vectors, being even under inversion and odd under time-reversal symmetry. Furthermore, time-reversal switches the circular polarization of the phonon mode, $u_R \rightarrow u_L^*$. Assuming a magnetic field applied along the Cartesian $z$-direction, chosen to be perpendicular to the phonon modes, the free energy, invariant under time-reversal and inversion symmetry, takes the form
\begin{multline}
  f(M_z, B_z, u_R, u_L; T) = f_M(T) + f_u(T) \\ - M_z B_z - \frac{1}{2} \chi B_z^2 - \alpha \left(B_z - \mu_0 M_z\right) (u_R u_R^*-u_L u_L^*).   
\end{multline}
Here, $f_M$ denotes the temperature-dependent free energy describing the magnet in the absence of an applied magnetic field (e.g. Phi-4 theory for Ising ferromagnet). Similarly, $f_u$ describes the free energy of the bare phononic part. The macroscopic magnetization is obtained from the free energy via the statistical physics relation $\mathcal{M} = -\frac{\partial f}{\partial B_z}$, and contains three terms: the spontaneous magnetization, the induced magnetization, and the phonomagnetic contribution
\begin{equation}
    \mathcal{M} = -\frac{\partial f}{\partial B_z} = \underbrace{M_z}_{\text{spontaneous}} + \underbrace{\chi B_z}_{\text{induced}} + \underbrace{\alpha (u_R u_R^*-u_L u_L^*)}_{\text{phonomagnetic}}.\label{free:energy}
\end{equation}
The magnetization induced by axial phonons (phonomagnetic contribution) is analogous to the optical inverse Faraday effect: both effects are caused by an imbalance between the amplitudes of right and left circularly polarized perturbation fields.
The phononic effect can therefore be seen as the phonon inverse Faraday effect. 
However, the key difference between these two effects is that the phonon inverse Faraday effect is caused by the lattice vibrations, while its optical counterpart is light-induced. While in recent experiments circularly polarized phonon modes were induced by a laser pump \cite{luo2023, Basini2024, Davies2024}, it is important to point out that phonon inverse Faraday is general and does not require light to occur. Instead, other excitation mechanisms, such as thermal gradients, would give a phenomenologically similar contribution. 

In the absence of an applied magnetic field ($B_z = 0$) and a spontaneous magnetization ($M_z = 0$), the magnetization is entirely determined by the phononic part
\begin{equation}\label{landau_theory_magnetization}
    \mathcal{M} = M_{\text{ph}} = \alpha (u_R u_R^*-u_L u_L^*),
\end{equation}
where $\alpha$ is a real-valued coefficient to be determined by a microscopic theory.

By analogy to the optical Faraday effect, and as the name suggests, a reciprocal effect to the phonon inverse Faraday effect exists, the phonon Faraday effect.
In the case of the optical Faraday effect, a linearly polarized laser beam rotates its polarization direction while passing through a magnetic medium. Similarly, linearly polarized phonons rotate their polarization direction while passing through a magnetic medium. This effect can be derived from the same free energy, equation \eqref{free:energy}. In analogy to the dielectric constant, we define the phonon response function, $\epsilon_{R,L} = \frac{\partial^2f}{\partial u_{R,L}\partial u^*_{R,L}}$. The part dependent on the applied magnetic field and spontaneous magnetization follows to be~\cite{shabala2024phonon} 
\begin{align}
    \Delta\epsilon^u_R &= -4\pi \frac{\partial^2F_u}{\partial u_R \partial u_R^*} = -4\pi \alpha \left(B_z - \mu_0 M_z\right), \\     \Delta\epsilon^u_L &= -4\pi \frac{\partial^2F_u}{\partial u_L \partial u_L^*}= \hphantom{-}4\pi \alpha \left(B_z - \mu_0 M_z\right).
    \label{eq:phononepsilon}
\end{align}
As a result, the phonon response function differs for left and right circularly polarized phonons, leading to a phonon birefringence in the magnetic sample and the phonon Faraday effect. While the correspondence between optical and phonon (inverse) Faraday effects was discussed in great detail here, we note that the paradigm of phonon analogues of optical effects is more general and also encompasses other examples, e.g., the phonon Cotton-Mouton effect \cite{Juraschek2020phono}. 

So far, the discussion has centered around magnetic effects arising due to optical phonons. However, we would like to point out that the phonon Faraday effect has also been observed for acoustic waves, as shown on the example of single-crystal yttrium-iron garnet and a pulse-echo method~\cite{Matthews1962}.

\section{Experiments and giant phono-magnetic effect}\label{section:experiments}
\begin{figure}
    \centering
    \includegraphics[width=0.49\textwidth]{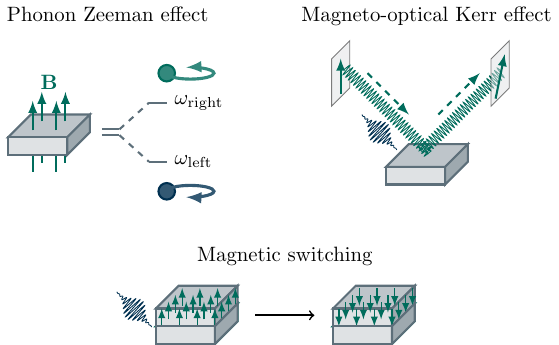}
    \caption{Visualization of the physical phenomena used to detect phono-magnetic effects: The Phonon Zeeman effect, i.e. splitting of a phonon mode into left and right circularly polarized phonons; the magneto-optical Kerr effect which constitutes polarization rotation of a probe field reflected off a magnetized sample; Magnetic switching where circularly polarized phonon modes induce switching of magnetic order in a structure placed on top of the substrate.}
    \label{fig:experiments}
\end{figure}

Phono-magnetic effects, particularly the giant dynamically induced magnetization due to axial phonons, have been observed in several experiments. Notably, these experiments were conducted by various independent groups using different measurement schemes and diverse classes of materials, underscoring the universality of phono-magnetism. The most prominent physical phenomena utilized in these experiments are illustrated in Figure \ref{fig:experiments}, including the phonon Zeeman effect, the magneto-optical Kerr effect, and magnetic switching. A summary is provided in Table \ref{tab:experiments}, with further details discussed subsequently.

\subsection{Phonon Zeeman effect}\label{subsec:exp_phonon_Zeeman}
We start with the experiments based on the phonon Zeeman effect, illustrated in Figure \ref{fig:experiments}.
This effect is characterized by the splitting of a two-fold degenerate phonon mode into two circularly polarized modes with opposite helicities in the presence of an applied magnetic field. The magnetic field $\vec{B}$ couples to the phonon mode through Zeeman coupling of the form $\vec{\mu}_{\text{ph}} \cdot \vec{B}$, where $\vec{\mu}_{\text{ph}}$ is the phonon magnetic moment~\cite{juraschek2019orbital}. Hence, measuring the frequency difference of left and right circularly polarized phonon modes, $\Delta\omega = \omega_{\text{R}} - \omega_{\text{L}}$, reveals the phonon magnetic moment at low field strengths~\cite{Anastassakis1972, Dohm1975,juraschek2017dynamical},
\begin{equation}
    \hbar \Delta\omega = \vec{\mu}_{\text{ph}} \cdot \vec{B}. 
    \label{eq:Zeeman}
\end{equation}

\begin{figure*}
    \centering
  \includegraphics[width=0.9\textwidth]{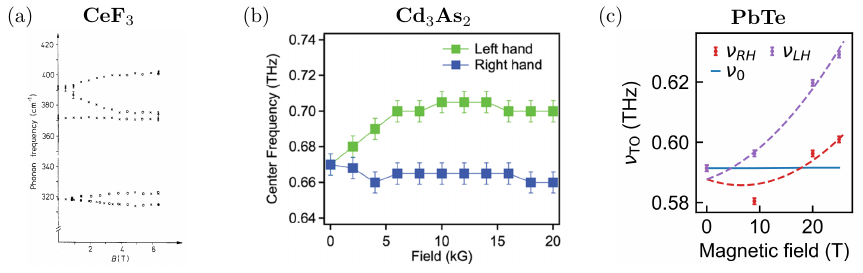}
    \caption{Phonon Zeeman splitting in CeF$_3$ observed using Raman scattering~\cite{Schaack1976}, in Cd$_3$As$_2$ observed using time-domain magnetoterahertz spectrometry \cite{cheng2020large}, and in PbTe observed with polarization-dependent terahertz spectroscopy \cite{baydin2022magnetic}. 
    \textit{(a, reproduced with permission from~\cite{Schaack1976}; Copyright IOP Publishing. All rights reserved.)}
    \textit{(b, reproduced with permission from~\cite{cheng2020large}; Copyright 2020 American Chemical Society)}
    \textit{(c, reproduced with permission from~\cite{baydin2022magnetic}; Copyright 2022 American Physical Society)}
    }
    \label{fig:phonon_Zeeman}
\end{figure*}

The first observations of the phonon Zeeman effect go back to the 1970s. Using Raman spectroscopy, Schaack recorded the splitting of phonon modes in CeF$_3$~\cite{Schaack1975, Schaack1976} and CeCl$_3$ at $\approx 2~K$ and magnetic field strengths of up to 6~T~\cite{Schaack1977}. 

In CeF$_3$, it has been observed that the E$_g$ phonons at 204~cm$^{-1}$ (6.1~THz) and 392~cm$^{-1}$ (11.75~THz) considerably split in an applied magnetic field. Low temperature experiments using Raman scattering~\cite{Schaack1975,Schaack1976} observe a respective splitting of 1.8 cm$^{-1}$ (0.05~THz) and 5.9 cm$^{-1}$ (0.18~THz) in a magnetic field of 1~T. Comparing these splittings with equation \eqref{eq:Zeeman} gives giant corresponding phonon magnetic momenta of $3.8~\mu_B$ and $12.6~\mu_B$. For high magnetic fields $B>3~\text{T}$, the phonon Zeeman splitting saturates to a constant value with a saturation splitting of 8~cm$^{-1}$ (0.24~THz) and 26.7~cm$^{-1}$ (0.8~THz), respectively. 

In CeCl$_3$, the 109~cm$^{-1}$ Raman active $E_{2g}$ mode shows a Zeeman splitting of roughly 4.4~cm$^{-1}$ (0.13~THz) at a magnetic field strength of 1~T \cite{Schaack1977}, translating to a phonon magnetic moment of $9.3\mu_B$ (see Figure~\ref{fig:phonon_Zeeman}~(a)). The saturation splitting for high magnetic fields is $7~\text{cm}^{-1}$ (0.2~THz). The effect is even stronger for the higher lying 197~cm$^{-1}$ $E_{1g}$ phonon mode, with a splitting of $\approx10$~cm$^{-1}$ (0.3~THz) at 1~T, and a corresponding phonon magnetic moment of 21~$\mu_B$. Here, the saturation splitting is 18~cm$^{-1}$ (0.5~THz).

Raman spectroscopy was also used in more recent experiments that resulted in similar findings. For example, in MoS$_2$ phonon Zeeman splitting of E'' mode at 270 cm$^{-1}$ (8.1 THz) indicates a phonon magnetic moment of approximately $2.5\mu_\text{B}$ \cite{tang2024exciton, mustafa2025origin}. 
This value is in agreement with the calculations performed with the recently proposed microscopic theory of the effect \cite{chaudhary2023giant}.

$E_g$ modes in CoTiO$_3$ also demonstrate a phonon Zeeman splitting of a similar size. 
In this material, for an external magnetic field of 7 T, the $E_g^{(1)}$ mode at roughly 200 cm$^{-1}$ (6 THz) demonstrates a splitting of about 7.14 cm$^{-1}$ (0.21 THz), while the $E_g^{(2)}$ mode at approximately 275 cm$^{-1}$ (8.3 THz) shows a splitting of roughly 2.03 cm$^{-1}$ (0.06 THz) \cite{lujan2024spin}.
This indicates a phonon magnetic moments of 1.11 and 0.29 $\mu_\text{B}$ respectively.

In Fe$_{1.75}$Zn$_{0.25}$Mo$_3$O$_8$, the phonon magnetic moment was observed to be $0.22$ $\mu_\text{B}$ for the P1 mode at 42 cm$^{-1}$ (1.26 THz). 
Notably, near the Néel temperature, the phonon magnetic moment reached values of $2.62 \, \mu_\text{B}$ \cite{wu2025magnetic}. 
However, another P1 phonon mode in Fe$_{1.75}$Zn$_{0.25}$Mo$_3$O$_8$ at 51 cm$^{-1}$ (1.53 THz) shows a smaller magnetic moment \cite{wu2025magnetic}.

In Fe$_2$Mo$_3$O$_8$, the splitting of the P1 phonon mode at 42 cm$^{-1}$ (1.26 THz) at 9 T indicated a phonon magnetic moment of 0.11 $\mu_\text{B}$ \cite{wu2023fluctuation}. Similarly to Fe$_{1.75}$Zn$_{0.25}$Mo$_3$O$_8$, near the Néel temperature the phonon magnetic moment was shown to increase drastically, reaching values of 0.68 $\mu_\text{B}$.

For the centrosymmetric ferromagnet Co$_3$Sn$_2$S$_2$, the phonon magnetic moment also demonstrates a dependence on the temperature. The phonon Zeeman splitting in this material reaches 1.27 cm$^{-1}$ (0.04 THz) at low temperatures, but decreases with increasing temperature, disappearing next to the Curie temperature \cite{che2025magnetic}. This measurement was performed on doubly degenerate $E_g$ phonon modes at around 295 cm$^{-1}$ (8.67 THz).
Additionally, this experiment shows that when an additional external magnetic field is applied, the phonon magnetic moment changes less than due to the internal magnetic field.
Thus, the intrinsic magnetic order of the ferromagnet induces chiral phonons.

The dependency of the phonon magnetic moment on Curie and Néel temperature demonstrated for Fe$_{1.75}$Zn$_{0.25}$Mo$_3$O$_8$, Fe$_2$Mo$_3$O$_8$ and Co$_3$Sn$_2$S$_2$ \cite{wu2025magnetic, wu2023fluctuation, che2025magnetic} highlights the dependency of axial phonons on the magnetic order. 
\begin{table}[]
    \centering
    \small
    \begin{tabular}{llll}
    \hline\hline
     Material & Mode & Mag. mom. ($\mu_B$) & Ref. \\
    \hline
     CeF$_3$ & E$_g$ (6.1~THz) & 3.8 & \cite{Schaack1975,Schaack1976} \\
     & E$_g$ (11.75~THz) & 12.6 & \cite{Schaack1975,Schaack1976} \\
    CeCl$_3$ & E$_{2g}$ (3.27~THz) & 9.3 & \cite{Schaack1977} \\
      & E$_{1g}$ (5.91~THz) & 21 & \cite{Schaack1977} \\
      MoS$_2$ & E$''$ (8.1~THz) & $ 2.4-2.5$ & \cite{tang2024exciton,mustafa2025origin} \\
     CoTiO$_3$ & E$_g^{(1)}$ (6~THz) & 1.11 & \cite{lujan2024spin} \\
     CoTiO$_3$ & E$_g^{(2)}$ (8.3~THz) & 0.29 & \cite{lujan2024spin} \\
     Fe$_{1.75}$Zn$_{0.25}$Mo$_{3}$O$_8$ & P1 1.26~THz & $2.62$ (Near $T_N$) & \cite{wu2025magnetic} \\
     Fe$_2$Mo$_{3}$O$_8$ & P1 (1.26~THz) & $0.68$ (Near $T_N$)& \cite{wu2023fluctuation} \\
     Co$_3$Sn$_{2}$S$_2$ & E$_g$ (8.67~THz) & & \cite{che2025magnetic} \\
     Cd$_3$As$_2$ & E$_u$ (0.67~THz) & $2.7$ & \cite{cheng2020large} \\
     PbTe & T$_{1u}$ (1.25~THz) & $ 4 \times 10^{-2}$ & \cite{baydin2022magnetic} \\
     Pb$_{0.4}$Sn$_{0.6}$Te & TO$_1$ (0.9~THz) & $1.2$ & \cite{hernandez2023observation} \\
     Pb$_{0.4}$Sn$_{0.6}$Te & TO$_2$ (1.6~THz) & $3.3$ & \cite{hernandez2023observation} \\
    \hline\hline
    \end{tabular}
    \caption{Summary of the experiments recording phonon magnetic moment through phonon Zeeman effect with the material, phonon mode which exhibits the Zeeman splitting and the observed magnetic moment.}
    \label{tab:experiments}
\end{table}

Also with Raman spectroscopy and Zeeman splitting, Ning et al. show that the hybridization between phonon and magnon modes in the antiferromagnetic FePSe$_3$ leads to spontaneous generation of elliptically polarized phonons \cite{ning2024spontaneous}.
This is additionally interesting because the original phonon mode is energetically nondegenerate. 
In most other cases where Zeeman splitting was used as a way to detect axial phonons, the degeneracy was a prerequisite for studying phono-magnetic effects experimentally.

The phonon Zeeman effect can also be detected using time-domain THz spectroscopy. 
Using this technique on PbTe allows recording a large Zeeman splitting of the doubly degenerate $T_{1u}$ phonon mode at 42 cm$^{-1}$ (1.25 THz), which implies a phonon magnetic moment of approximately $4 \times 10^{-2} \mu_\text{B}$ \cite{baydin2022magnetic}. The splitting is depicted in Figure~\ref{fig:phonon_Zeeman}~(c).

Similarly, in the Dirac semimetal Cd$_3$As$_2$, time-domain THz spectroscopy indicates a phonon magnetic moment of 2.7 $\mu_\text{B}$~\cite{cheng2020large}, which is deduced by recording a splitting of the $E_u$ phonon mode at 0.67 THz. Figure~\ref{fig:phonon_Zeeman}~(b) shows the measurement data.

These results are on the same order of magnitude as the theoretical predictions for the same material made by Chen et al. based on emergent gauge theory \cite{Chen2025gauge}.

In the topological crystalline insulator Pb$_{0.4}$Sn$_{0.6}$Te, the phonon magnetic moment is observed through the splitting of TO$_1$ and TO$_2$ modes at 30 cm$^{-1}$ (0.9 THz) and 53 cm$^{-1}$ (1.6 THz), respectively.
In particular, the experiment on Pb$_{0.4}$Sn$_{0.6}$Te \cite{hernandez2023observation} shows that the phonon magnetic moment depends on the topological phases of the material.
Pb$_{1-x}$Sn$_{x}$Te is a topological crystalline insulator for $x>0.32$ and with a trivial phase for $x\leq0.32$. 
The phonon magnetic moment increased by two orders of magnitude during the transition from trivial to topological phase, reaching values of 1.2 $\mu_\text{B}$ for TO$_1$ mode and 3.3 $\mu_\text{B}$.
This dependency indicates a relationship between topology and phono-magnetic effects.
\begin{figure*}
    \centering
\includegraphics[width=0.98\linewidth]{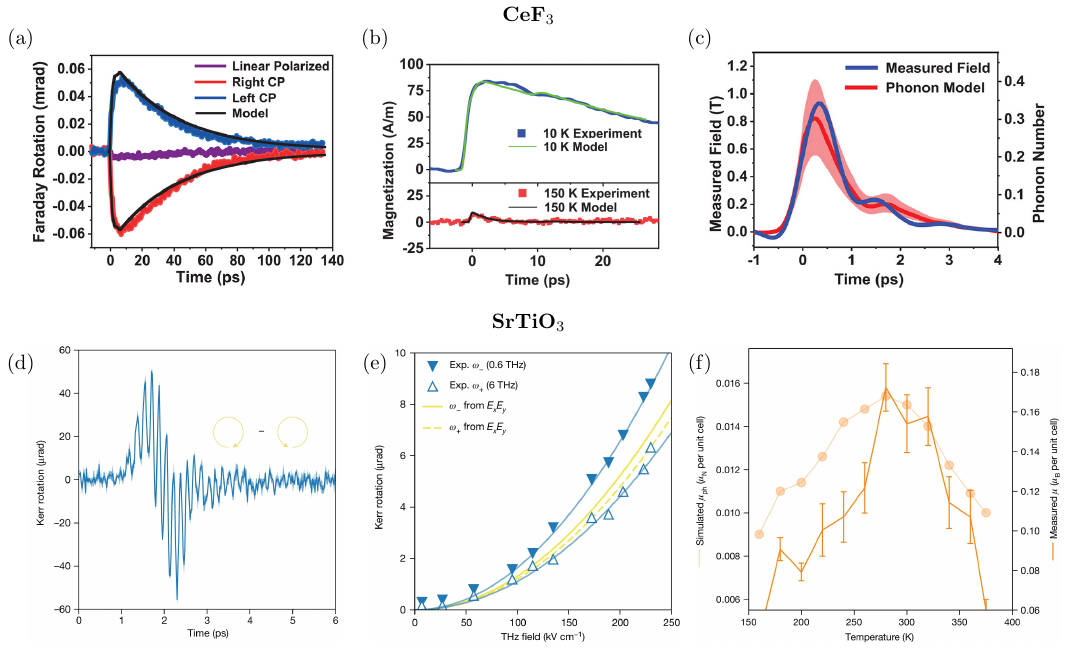}
    \caption{Dynamic magnetization induced by axial phonons, measured using the magneto-optical Kerr effect (MOKE). (a)-(c) show results for CeF$_3$ \cite{luo2023}, (d)-(e) show results for SrTiO$_3$~\cite{Basini2024}. (a) Faraday rotation in CeF$_3$ at 10~K after excitation with circularly polarized THz pump. (b) Corresponding sample magnetization computed from (a) using the temperature-dependent Verdet constant of CeF$_3$. (c) Deduced effective magnetic field due to axial phonons, giving rise to the magnetization in (a) and (b). (d) Faraday rotation in SrTiO$_3$ at room temperature after excitation with circularly polarized THz pump. (e) Quadratic dependence of the Kerr rotation (magnetization) on the field strength of the THz laser pulse. (f) Comparison of theoretical and experimental phonon magnetic moment. \textit{(a)-(c), reproduced with permission from~\cite{luo2023}; Copyright 2023 The Authors, some rights reserved. }\textit{(d)-(f), reproduced from~\cite{Basini2024} under the Creative Commons license. }}
    \label{fig:MOKE}
\end{figure*}

\subsection{Measurements based on the magneto-optical Kerr Effect}

Another way of measuring phonon-induced magnetization is through the magneto-optical Kerr effect (MOKE), which constitutes a rotation of polarization of linearly polarized light when reflected from a magnetized surface \cite{Freiser1968}. MOKE is related to the Faraday effect, where the rotation of polarization occurs as a result of transmission through a magnetized sample.
Because of the linear relationship between the magnetic field induced by the sample and the Kerr angle, measuring the Kerr angle, i.e., the rotation of polarization, provides a way to estimate the material's magnetization.

Luo \textit{et al.}~\cite{luo2023} performed pump-probe MOKE experiments on the rare-earth halide CeF$_3$. Complementary to the phonon Zeeman experiments by Schaack~\cite{Schaack1975, Schaack1976, Schaack1977} mentioned above and measuring the $E_g$ Raman modes, the experiments by Luo \textit{et al.} used a circularly polarized THz pump laser to excite the infrared active E$_u$ mode at around 10.5~THz. The used laser pulse had an electric field strength of 560~kV/cm, with a pulse duration of 0.45~ps and a frequency of 10.8~THz (1~THz bandwidth). At low temperatures (10~K), the sample is ferromagnetic, with a long spin relaxation time of $\approx 40~\text{ps}$~\cite{luo2023}. In contrast, the lifetime of the phonons is much shorter, and only about $\approx 0.6~\text{ps}$. The long spin relaxation time can be clearly seen in the Faraday rotation of \autoref{fig:MOKE}(a) and the corresponding magnetization, obtained from the Faraday rotation and the temperature-dependent Verdet constant, \autoref{fig:MOKE}(b). This allows deducing the necessary pseudo-magnetic field induced by the axial phonon, which reaches an order of $\approx 1~\text{T}$, as shown in \autoref{fig:MOKE}(c). However, the peak measured magnetization, shown in \autoref{fig:MOKE}(b), is given by $\approx 85~\text{A/m}$. Given a primitive unit cell size of $\approx 315~\text{\AA}^3$ (space group P$\overline{3}$c1), the phonon magnetization per cell is tiny, $0.003~\mu_B \approx 5~\mu_N$.

A similar experiment was performed by Basini \textit{et al.}, measuring the phonon-induced magnetization in SrTiO$_3$~\cite{Basini2024}. In their setup, they pumped the ferroelectric soft mode, which has a frequency of $\approx 2.7~\text{THz}$ at room temperature. Basini \textit{et al.} used varying laser field strengths of 200-300~kV/cm. In contrast to CeF$_3$, SrTiO$_3$ is a nonmagnetic (quantum) paraelectric material. As a result, the dynamically induced magnetization decays with the phonon lifetime, as can be seen in the Kerr rotation measurement shown in \autoref{fig:MOKE}(d), comparing the signals obtained for left- and right-circularly polarized pump pulses. An important experimental signature in their measurements is the square dependence of the induced magnetization, measured by the Kerr angle, against the pump laser field strength, as shown in \autoref{fig:MOKE}(e), consistent with equations \eqref{eq:phonomag} and \eqref{eq:phonoeqm}. Comparing the measured Faraday rotation with the dynamical Verdet constant in SrTiO$_3$, one obtains an effective magnetic field strength of 35~mT, which is about 3 orders of magnitude lower, as compared to CeF$_3$~\cite{Basini2024}. Basini \textit{et al.} translate this effective magnetic field to a large phonon magnetic moment of $\approx 0.1~\mu_B$~\cite{Basini2024}, an interpretation which is opposed in Refs~\cite{merlin2023unraveling, merlin2025magnetophononics}.

\subsection{Proximity-induced magnetization switching}
The induction of pseudomagnetic fields by axial phonons opens the perspective for proximity-induced manipulation of the magnetic properties of a material.
The experimental evidence for such a possibility was recently reported by Davies et al. \cite{Davies2024}, who used a setup consisting of substrates of Al$_2$O$_3$, SiO$_2$ or Si with a heterostructure composed of GdFeCo and Si$_3$N$_4$ placed on top of it.
Axial phonons were induced in the substrate through driving by a circularly polarized laser field, resulting in the axial phonon induced, pseudomagentic field. This pseudomagnetic field causes a switching of the magnetic order of the heterostructure placed on top of the substrate~\cite{Davies2024} as illustrated in Figure~\ref{fig:experiments}. Interestingly, even though the magnetization induced by axial phonons is short-lived, the switching of the magnetic order is permanent.

\section{Microscopic theories}\label{sec:micro-theories}

In the previous sections, we gave an overview of the phenomenological arguments for the emergence of magnetic effects from axial phonons. 
We have also discussed experimental evidence for such effects.
In this section, we will summarize recently proposed microscopic theories of phonon-induced magnetism. 
It is worth noting that what unifies all of these theories is the consideration of the role of electrons in these effects.

The relevance of electrons in phono-magnetic effects can be motivated by considering the gyromagnetic ratio $\gamma$, which determines the magnetic momentum of a particle with an angular momentum: $\vec{\mu} = \gamma \vec{L}$. The gyromagnetic ratio is determined by the charge $q$ and mass of the particles $m$. 
Classically, for a rotating body, it is given by $\gamma = \frac{q}{2m}$.
Thus, for an electron with spin and orbital angular momentum, the magnetic moment is given by the Bohr magneton, $\mu_{\text{B}} = \frac{e\hbar}{2m_{\text{e}}}$. 
At the same time, if we consider the magnetic moment of the phonon, assuming that it is determined only by the angular momentum of the phonon, it should be given by $\mu_{\text{ph}} =\frac{e\hbar}{2m_{\text{ion}}}$. 
This implies that the effect should be a factor of $\frac{m_\text{ion}}{m_{\text{e}}}$ smaller than the Bohr magneton \cite{geilhufe2023KTO}. 
At the same time, as described in section \ref{section:experiments}, the experimental observations suggest a phonon magnetic moment on the scale of the Bohr magneton, which is determined by the mass of electrons.

Additionally, the Einstein-de Haas effect and the Barnett effect hint at the role of electrons in phonon-induced magnetism. 
These effects are reciprocal to each other and connect mechanical rotation to magnetization. 
The Einstein-de Haas effect refers to the rotation of the body caused by the change in its magnetic moment, while the Barnett effect describes the opposite: magnetization induced by the change in angular momentum.

The relevance of phonon angular momentum in the Einstein-de Haas effect was first pointed out by Zhang and Niu, who showed that when calculating the change of the electron angular momentum in the Einstein-de Haas effect, phonon angular momentum needs to be considered \cite{zhang2014angular}. Similarly, recent experiments on phono-magnetic effects \cite{Davies2024, Basini2024} explain their findings in terms of the phonon Barnett effect. 
For example, Basini et al. explain their measurements of the magnetic moment on the scale of the Bohr magneton by phonon angular momentum being transferred to the total angular momentum of electrons, thus inducing stronger magnetization.

Evidence for angular momentum transfer between electrons and phonons was also recorded by Tauchert et al. \cite{Tauchert2022}. 
In their observation of ultrafast demagnetization of a material through the Einstein-de Haas effect, Tauchert et al. recorded an intermediate stage between a magnetized non-rotating and demagnetized rotating stages. The intermediate state is characterized by the presence of circularly polarized phonons in the material, which shows angular momentum transfer between electrons in the magnetized stage and phonons, resulting in the circular motion of ions.

\begin{figure}
    \centering
    \includegraphics[width=0.85\linewidth]{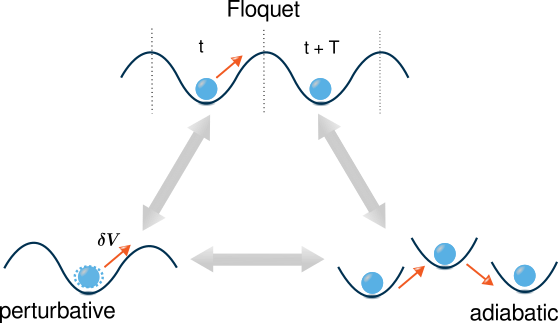}
    \caption{Schematic overview of the three categories of microscopic theories (perturbative, adiabatic, Floquet), which are limiting cases of each other.}
    \label{fig:approach_triangle}
\end{figure}

In the following, we summarize recently proposed theoretical approaches that examine the role of electrons in phono-magnetic effects.
As we will describe below, these theories can be grouped into three larger categories: perturbative, adiabatic and Floquet approaches. Perturbative approaches assume that the interaction of electrons and phonons is weak, adiabatic approaches assume that the phonon dynamics is much slower than the electron dynamics, and Floquet approaches approximate the phonon as a continuous wave. As depicted in \autoref{fig:approach_triangle}, these three approaches are limiting cases of each other. 
As detailed later, the adiabatic description becomes perturbative when considering the low-frequency limit.
Both the Floquet and the adiabatic approach connect the phono-magnetic effect to the geometric phase.
And the perturbative approach is obtained from the Floquet approach by considering small changes. 
The underlying assumptions and approximations are valid in different materials, as the next sections will show, but all consider a coupling between electrons and phonons.

\subsection{Perturbative approaches}
A first class of microscopic theories considers the interaction between phonons and electrons as a perturbation to the system.

\subsubsection{Electron-phonon coupling}
In the Born-Oppenheimer approximation, the total Hamiltonian of electrons in the crystalline potential can be written as
\begin{equation}
    \op{H} = \op{H}_e + \op{H}_p + \op{H}_{ei}.
\end{equation}
Here, $\op{H}_e$ denotes the pure electronic part, $\op{H}_p$ the phononic part, and $\op{H}_{ei}$ the attractive interaction between ions and electrons
\begin{align}
    \op{H}_e &= \sum_i \left[\frac{\op{\vec{p}}_i^2}{2m} + U_{el}(\left|\vec{r}_i - \vec{r}_j\right|)\right], \\
    \op{H}_p &= \sum_{\vec{k}\nu} \hbar\omega_{\vec{k}\nu} \op{a}^\dagger_{\vec{k}\nu} \op{a}_{\vec{k}\nu}, \\
    \op{H}_{ei} &= \sum_{i}\sum_{l\alpha} U(\left|\vec{r}_i - \vec{R}_{l\alpha}\right|).
\end{align}
To connect the ionic potential $\op{H}_{ei}$ to the phononic sector, one expands the potential $U$ by assuming small lattice displacements, according to equation \eqref{eq:disp},
\begin{equation}
    \op{H}_{ei} \approx \sum_{i}\sum_{l\alpha} \left[U(\left|\vec{r}_i - \vec{R}^{(0)}_{l\alpha}\right|)+ \left(\nabla_{\vec{\tau}_{l\alpha}} U\right)\cdot \vec{\tau}_{l\alpha} + \dots \right].\label{eq:elphon}
\end{equation}
To first order, this gives rise to an interaction written in the displacement coordinate $\vec{\tau}_{l\alpha}$, i.e., $\left(\nabla_{\vec{\tau}_{l\alpha}} U\right)\cdot \vec{\tau}_{l\alpha}$, mediating the electron phonon interaction. 

\subsubsection{Phonon inverse Faraday effect}\label{subsec:phononIFE}
The phonon inverse Faraday effect~\cite{shabala2024phonon, Juraschek2020phono, juraschek2017dynamical, rebane1983faraday} describes the induction of a DC magnetization due to circularly polarized (axial) phonons (see Section \ref{subsec:Landau_theory}). The microscopic theory of the phonon inverse Faraday effect was derived by Shabala and Geilhufe~\cite{shabala2024phonon} and is based on time-dependent second-order perturbation theory, similar to its optical analogue~\cite{pershan1966theoretical}. 

For a general time-dependent perturbation,
\begin{equation}
    V(t) = v(t) e^{\mathrm{i}\omega t} + v^*(t) e^{-\mathrm{i}\omega t},
\end{equation}
second-order time-dependent perturbation theory yields an effective Hamiltonian of the form \cite{pershan1966theoretical, Wong2025}
\begin{equation}
    \mathcal{H}^{ab}_{\text{eff}} = - \sum_n \left[\frac{\bra{a}v\ket{n}\bra{n} v^*\ket{b}}{E_{nb}-\hbar\omega} + \frac{\bra{a}v^*\ket{n}\bra{n} v\ket{b}}{E_{nb}+\hbar\omega}\right],
    \label{eq:scndorder}
\end{equation}
which describes the mixing of two states $a$ and $b$ via virtual transitions involving the states $n$. For a dynamic displacement $\tau_{l\alpha}$, we can use Equation \eqref{eq:elphon} to identify the dynamical perturbation $v$ as follows:
\begin{equation}
    v(t) = \sum_{l\alpha} \left(\nabla_{\vec{\tau}_{l\alpha}} U\right) \cdot \vec{\tau}_{l\alpha}.
\end{equation}
In general, this allows us to express the effective Hamiltonian in terms of a generalized susceptibility:
\begin{equation}
    \mathcal{H}^{ab}_{\text{eff}} = - \chi^{ab}_{ij} \tau_{l\alpha i} \tau_{l\alpha j}.
\end{equation}
As we are interested in the induction of magnetization due to an axial phonon, we focus on the time-reversal symmetry-breaking contribution in the effective Hamiltonian and disregard all other terms:
\begin{multline}
  \mathcal{H}^{ab}_{\text{eff}} = -\hbar \omega \sum_{l\alpha}\left[\vphantom{\frac{\left[\left(\nabla_{\vec{\tau}_{l\alpha}} U\right)^*_{nb}\right]_z}{E_{nb}^2}} \left(\vec{\tau}_{l\alpha}\times \vec{\tau}^*_{l\alpha}\right)_z \times \right. \\ \left. \times \sum_n \frac{\left[\left(\nabla_{\vec{\tau}_{l\alpha}} U\right)_{an} \times \left(\nabla_{\vec{\tau}_{l\alpha}} U\right)^*_{nb}\right]_z}{E_{nb}^2-\hbar^2\omega^2}\right] \\+ \text{other 2$^{\text{nd}}$ order terms}.\label{eq:effHamIFE}
\end{multline}
Here, we used the abbreviation $\left(\nabla_{\vec{\tau}_{l\alpha}} U\right)_{an} = \bra{a}\nabla_{\vec{\tau}_{l\alpha}} U\ket{n}$.
A similar expression was independently obtained in Ref. \cite{merlin2025magnetophononics}, where electron–phonon coupling was also treated as a time-dependent perturbation. 
In that work, the quantity $\left(\vec{\tau}_{l\alpha}\times \vec{\tau}^*_{l\alpha}\right)_z$ was identified as a time reversal symmetry breaking field. Such a field can be interpreted as a non-Maxwellian effective magnetic field, i.e. a field that reproduces the effects of a real magnetic field inside the material but remains undetectable outside the sample \cite{merlin2023unraveling, merlin2025magnetophononics}.

Equation \eqref{eq:effHamIFE} can be brought into the form where electron-phonon coupling is expressed in terms of electron-phonon matrix elements.
To do that, we first use the fact that in a periodic solid, the states $\ket{n}$ are Bloch states of the form $\ket{\vec{k}n}$. 
Transforming to mode coordinates allows us to express the gradient terms using electron-phonon matrix elements~\cite{Giustino2017},
\begin{multline}\label{matrix_elements}
    g_{mn\nu}(\vec{k}, \vec{q}) = \bra{m, \vec{k}+\vec{q}} \sum_{l} l_{\vec{q}\nu} e^{i\vec{q}\cdot\vec{R_l}} \frac{\partial U}{\partial \tau_{l\nu}} \ket{n,\vec{k}}.
\end{multline}
Imposing the quantization of the phonon displacement according to Equation \eqref{eq:quantizeddisp} gives the following form of the effective Hamiltonian:
\begin{equation}
    \mathcal{H}_{\text{eff}}^{ab}(\vec{k}) = -i \sum_{\vec{q}}(\hat{a}_{\vec{q},\mu} + \hat{a}_{-\vec{q},\mu}^\dagger)(\hat{a}_{-\vec{q},\nu}^{\dagger}+\hat{a}_{\vec{q},\nu})\Pi^{ab}_{\mu\nu}(\vec{q},\vec{k}),
    \label{eq:effHamIFEquant}
\end{equation}
with
\begin{multline}
    \Pi^{ab}_{\mu\nu}(\vec{q},\vec{k}) = \hbar \omega\sum_n\left[ \frac{g_{an\mu}(\vec{k},\vec{q})g_{bn\nu}^*(\vec{k},\vec{q})}{E_{\vec{k}nb}^2-\hbar^2\omega^2} \right.\\ \left.-\frac{g_{an\nu}(\vec{k},\vec{q})g_{bn\mu}^*(\vec{k},\vec{q})}{E_{\vec{k}nb}^2-\hbar^2\omega^2}\right]\label{eq:Pi}
\end{multline}
Equations \eqref{eq:effHamIFE} and \eqref{eq:effHamIFEquant} have several important consequences. First, the ionic gradient of the potential $\nabla_{\vec{\tau}_{l\alpha}} U$ is odd under spatial inversion. Hence, for a system with broken inversion symmetry, matrix elements of the form $\bra{a}v\ket{n}$ generally exist. In contrast, for systems with inversion symmetry, non-zero contributions require states $\ket{a}$ and $\ket{n}$ with opposite parity. In materials like SrTiO$_3$, this would involve, for example, a virtual transition between oxygen $p$-states and titanium $d$-states (see Figure~\ref{fig:chaudhary_etal}~(b)). Second, the denominator contains the electronic energy difference $E_{nb} = E_n - E_b$. In insulating materials, this difference is primarily dominated by the band gap, which can reach the order of several eV, whereas the phonon energy is on the order of several meV.

Second-order perturbation theory was applied to estimate the pseudomagnetic field in SrTiO$_3$~\cite{shabala2024phonon} by discussing the splitting of the oxygen $p_+$ and $p_-$ states, similar to an orbital Zeeman effect (see Figure~\ref{fig:chaudhary_etal}~(c)). Due to selection rules between even and odd orbitals, the electronic splitting between $p_+$ and $p_-$ states due to virtual transitions with the titanium $d$ states is given by
\begin{equation}
    \Delta E = \hbar\omega \frac{\abs{g}^2}{\Delta^2 - \hbar^2\omega^2} \left(n_0 + \frac{1}{2}\right).
\end{equation}
This equation is dominated by the band gap of SrTiO$_3$, which is $\Delta = 3.75~\text{eV}$. In contrast, the typical electron phonon coupling strength is weaker by almost three orders of magnitude~\cite{Gastiasoro2023, Zhou2018}. The phonon energy $\hbar \omega$ for the ferroelectric soft mode of SrTiO$_3$ is $\approx 11~\text{meV}$~(2.7~THz).

\subsubsection{Orbit-lattice coupling of localized electrons}\label{subsec:SOC}

\begin{figure}
    \centering
    \includegraphics[width=\linewidth]{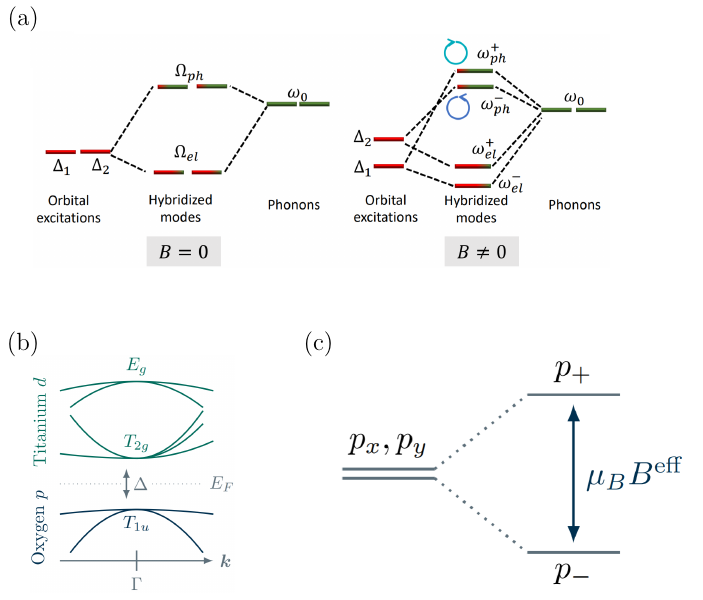}
    \caption{Emergence of chiral phonon modes from orbit lattice coupling. 
    (a) Modified frequency modes $\Omega_{el}$ and $\Omega_{ph}$ emerging from electron-phonon coupling with and without external magnetic field $B$. When an external magnetic field is present, time-reversal symmetry is broken and the degeneracy of the electronic states is lifted, leading to the formation of chiral hybridized modes $\omega_{ph}^{\pm}$.
    (b) The electronic structure of SrTiO$_3$ with the valence band consisting of oxygen $p$-states and the conduction band of Ti-$d$ states. The large band gap, which is characteristic for SrTiO$_3$, is denoted by $\Delta$.
    (c) The electronic $p$-orbitals split and their gap is proportional to the effective magnetic field arising from the phonons.
    \textit{(a) reproduced with permission from~\cite{chaudhary2023giant}; Copyright 2024 American Physical Society.(b) and (c) reproduced from~\cite{shabala2024phonon} under the Creative Commons Attribution license.}}
    \label{fig:chaudhary_etal}
\end{figure}

In materials with strongly localized electrons, such as the $f$-electron systems CeF$_3$ and CeCl$_3$, equations \eqref{eq:scndorder} and \eqref{eq:effHamIFEquant} can be evaluated for a few electronic levels, split by the local crystal field. Such a system was discussed by Chaudhari \textit{et al.}~\cite{chaudhary2023giant}, considering two Kramers doublets, with states
\begin{align}
    \ket{\Psi_1} &= \ket{J_\alpha; m_\alpha} \\
    \ket{\Psi_2} &= \ket{J_\alpha; -m_\alpha} \\ \ket{\Psi_3} &= \ket{J_\beta; m_\beta} \\ 
    \ket{\Psi_4} &= \ket{J_\beta; -m_\beta}
\end{align}
For degenerate phonon modes $\mu$ and $\nu$, the electron phonon coupling can be written as follows,
\begin{equation}
    \op{H}_{\text{el-ph}} = \left(\op{a}^\dagger_\mu+\op{a}_\mu\right)\op{O}_\mu + \left(\op{a}^\dagger_\nu+\op{a}_\nu\right)\op{O}_\nu,
\end{equation}
with the operators
\begin{align}
    \op{O}_\mu &= g_\mu \ket{1}\bra{3} - g_\mu^*\ket{2}\bra{4} + \text{h.c.}, \\
    \op{O}_\nu &= g_\nu \ket{1}\bra{3} - g_\nu^*\ket{2}\bra{4} + \text{h.c.} .
\end{align}
In contrast to Chaudhari \textit{et al.}~\cite{chaudhary2023giant}, who used a Green function approach, we are going to formulate the effect of localized electrons on the phonon modes in terms of the renormalized phonon Hamiltonian, in line with the previous section,
\begin{equation}
    \op{H}_{\text{ph}} = \left(\begin{array}{c}
     \op{a}^\dagger_\mu  \\
     \op{a}^\dagger_\nu
    \end{array}\right)\left[
    \left(
    \begin{array}{cc}
        \hbar\omega & 0 \\
        0 & \hbar\omega
    \end{array}
    \right) +     \left(
    \begin{array}{cc}
        \Pi_{\mu\mu} & \Pi_{\mu\nu} \\
        \Pi_{\mu\nu}^* & \Pi_{\nu\nu}
    \end{array}
    \right)\right]
    \left(\begin{array}{c}
     \op{a}_\mu  \\
     \op{a}_\nu
    \end{array}\right).
    \label{eq:Chaud_phon}
\end{equation}
From equation \eqref{eq:Pi}, we find
\begin{align}
    \Pi_{\mu\nu} &= \sum_{ij} \Pi^{ij}_{\mu\nu} \\
    &= \hbar\omega \left[\frac{g_\mu g_\nu^*-g_\nu g_\mu^*}{E_{13}^2-\hbar^2\omega^2} + \frac{g_\mu g_\nu^*-g_\nu g_\mu^*}{E_{24}^2-\hbar^2\omega^2}  \right].
\end{align}
The effective Hamiltonian \eqref{eq:scndorder} also gives rise to diagonal terms $\Pi_{\mu\mu}$ and $\Pi_{\nu\nu}$~\cite{pershan1966theoretical,chaudhary2023giant}, which are given by
\begin{equation}
    \Pi_{\mu\mu} = \abs{g_\mu}^2\left(\frac{E_{13}}{E_{13}^2-\hbar^2\omega^2}+\frac{E_{24}}{E_{24}^2-\hbar^2\omega^2}\right),
\end{equation}
and a similar expression for $\Pi_{\nu\nu}$. Introducing the abbreviation
\begin{align}
    \Pi_\pm = \frac{1}{2}\left(\Pi_{\mu\mu}\pm\Pi_{\nu\nu}\right),
\end{align}
allows us to diagonalize the Hamiltonian in equation \eqref{eq:Chaud_phon}, to give renormalized phonon frequencies
\begin{equation}
\hbar \omega_\pm = \hbar\omega + \Pi_+ \pm \sqrt{\abs{\Pi_{\mu\nu}}^2+\Pi_-^2}.    
\label{eq:phononom}
\end{equation}
Applying a magnetic field will break the time-reversal symmetry in the Kramers doublets $\ket{\Psi_i}$ (see Figure~\ref{fig:chaudhary_etal}(a)). This change affects the phonon modes according to equation \eqref{eq:phononom}. Comparison with the phonon Zeeman effect given in equation \eqref{eq:Zeeman}, allows us to define the phonon magnetic moment by 
\begin{equation}
    \mu_{ph} = \frac{\hbar}{2}\frac{\partial(\omega_{ph}^{+}-\omega_{ph}^{-})}{\partial B}\Bigg|_{B\rightarrow0}.
\end{equation}
Using their model Chaudhary et al. calculate phonon magnetic moment in CeCl$_3$ and CoTiO$_3$  of the order of magnitude of 0.1 $\mu_B$.

\subsubsection{Spin structure from generalized electron phonon interaction}\label{subsec:perturbative_fransson}
An alternative approach for describing the influence of phonon angular momentum to the electronic spin was proposed by Fransson \cite{fransson2023chiral}, using a generalized electron phonon interaction
\begin{equation}
    \op{H}_{\text{e-ph}} = \sum_{\vec{k}\vec{q}\mu} \psi^\dagger_{\vec{p}+\vec{k}} \mat{U}_{\vec{k}\vec{q}\mu} \psi_{\vec{k}}\left(\op{a}_{\vec{q}\mu}+\op{a}^\dagger_{-\vec{q}\mu}\right).
\end{equation}
Here, the matrix $\mat{U}_{\vec{k}\vec{q}\mu} = U_{\vec{k}\vec{q}\mu} \sigma_0 + \vec{J}_{\vec{k}\vec{q}\mu}\cdot\vec{\sigma}$ is composed of a conventional electron-phonon coupling term, $U_{\vec{k}\vec{q}\mu} \sigma_0$ which does not couple to the electron spin, as well as a spin-dependent electron-phonon coupling term $\vec{J}_{\vec{k}\vec{q}\mu}\cdot\vec{\sigma}$. The spin dependent electron phonon interaction can be derived from the spin-orbit interaction,
\begin{equation}
    \op{H}_{\text{SOC}} = \frac{\xi}{2} \left[\vec{E}\times\vec{p} + \vec{p}\times\vec{E}\right]\cdot\vec{\sigma}.
\end{equation}
The electric field $\vec{E}$ experienced by the electron results from the binding potential $V$, by $\vec{E}=-\nabla V$. Finally, by expanding the binding potential in terms of small lattice displacements, similar to equation \eqref{eq:elphon}, it is possible to derive an expression for the spin-dependen electron phonon interaction~\cite{fransson2023chiral},
\begin{equation}
    \vec{J}_{\vec{k}\vec{q}\mu} = -\iu \xi \mat{U}_{\vec{k}\vec{q}\mu} \vec{k}\times\vec{q},
\end{equation}
with $\mat{U}_{\vec{k}\vec{q}\mu}$ being the conventional, spin-conservative electron-phonon interaction~\cite{Giustino2017}. This spin-dependent type of electron phonon interaction can be analyzed perturbatively, e.g., using Green function methods as done in Ref.~\cite{fransson2023chiral} to show that the angular momentum carried by axial phonons is transferred to the electron spin, inducing nontrivial spin texture and circulating spin currents. Furthermore, it can be shown that the axial phonons coupled to magnons restores the ferromagnetic order and enhances its stability by increasing the anisotropy energy for magnon excitations~\cite{fransson2025chiral}. This principle can be generalized, by describing the formation of axial Raman modes due to magnetic quadrupole order \cite{Sutcliffe2025}.

\subsection{Adiabatic approach}
The adiabatic evolution of a quantum state describes the limit where the time-dependence is slow enough to leave the system in its instantaneous eigenstate, dependent on a slowly varying parameter. 
Ionic motion in solids represents such an example where the typical timescale of picoseconds (meV) can be seen as sufficiently slow in comparison to the electronic timescales, typically being in the range of femtoseconds (eV). 
The adiabatic evolution of quantum states is also tightly connected to the emergence of the geometric or Berry phase~\cite{Berry1984,aharonov1987phase}. 
Furthermore, the Berry connection and Berry curvature play a significant role in the modern theory of polarization~\cite{King-smith1993,Resta1994,Spaldin2012} and magnetization~\cite{resta2010electrical}.

\subsubsection{Berry phase argument}\label{subsubsec:Berry_phase_argument}

The effect of the axial phonons on the electronic states can also be examined by considering their evolution in the adiabatic regime.
Starting with the Schrödinger equation, $H(t)\ket{\Psi} = i \hbar \partial_t \ket{\Psi}$, we assume a system with 
a degenerate energy spectrum. 
For such a system the solutions to the Schrödinger equation are of the form \cite{rigolin2010adiabatic, rigolin2012adiabatic}
\begin{equation}\label{eq:berry_psi}
    \ket{\Psi(t)} = \sum_n \sum_{g_n} e^{-i\omega_n t} \text{U}^n_{h_n g_n} \ket{n^{g_n}(t)},
\end{equation}
where it is assumed that at the beginning of the time evolution the system is in a ground state.
In equation \eqref{eq:berry_psi} indices $g_n$ and $h_n$ denote the degeneracy states and $\text{U}^n_{h_n g_n}$ refers to the matrix elements of a unitary matrix such that $\text{U}^n_{h_n g_n} = \mathcal{T} \exp{\left[-\int_0^t \bra{g_n}\partial_{t'} \ket{h_n}dt'\right]}$. 
Here $\mathcal{T}$ referres to the time ordering operator.

Let us now consider a Hamiltonian influenced by the phonon displacement, $H_{\vec{u}}(t)$.
We will further limit the discussion to the degenerate ground states. 
Then the solutions of the Schrödinger equation $H_{\vec{u}}(t)\ket{\Psi} = i \hbar \partial_t \ket{\Psi}$ can be written as:
\begin{equation}\label{eq:berry_psi_mn}
    \ket{\Psi(t)} = \sum_n \sum_{g_n} e^{-i\omega_n t} e^{-i \gamma_{mn}(t)}.
\end{equation}
Here, $\omega_n t$ is the dynamical phase and $\gamma_{mn}(t)$ is the geometric phase, that can be expressed as:
\begin{equation}
    \gamma_{mn}(t) = -\iu\int_0^t \bra{m_{\vec{u}, \vec{k}}}\partial_{t'} \ket{n_{\vec{u}, \vec{k}}}dt'.
\end{equation}
Here, $m_{\vec{u}, \vec{k}}$ $n_{\vec{u}, \vec{k}}$ refer to different degeneracies of the ground state.
To relate this result to axial phonons, we perform the transformation $\frac{\partial }{\partial t} = \frac{\partial}{\partial\vec{u}}\frac{\partial \vec{u}}{\partial t}$.
Then, the geometric phase becomes:
\begin{equation}
    \gamma_{mn}(t) =-\iu \int_0^t \mathrm{d}t \bra{m_{\vec{k};\vec{u}}}\nabla_{\vec{u}} \ket{n_{\vec{k};\vec{u}}} \dot{\vec{u}}.
\end{equation}
Here we can recognize the gauge field $A^{mn} = \bra{m_{\vec{u}, \vec{k}}}\nabla_{\vec{u}} \ket{n_{\vec{u}, \vec{k}}}$ \cite{wilczek1984appearence} and linearize it with respect a small displacement $\vec{u}$:
\begin{equation}
    \gamma_{mn}(t) \approx \iu \sum_{ij} \int_0^t \mathrm{d}t\, \partial_{u_i} A^{mn}_{u_i} \dot{u}_i u_j.
\end{equation}
The sum in the equation above can be decomposed into symmetric and antisymmetric parts. 
Similarly to the phonon inverse Faraday effect formalim discussed in section \ref{subsec:phononIFE}, we focus on the time reversal symmetry breaking terms, i.e. the antisymmetric part. 
Thus, for a circular ionic motion in $xy$-plane we can write the geometric phase as:
\begin{equation}
\begin{split}
    \gamma_{mn}(t) & = \iu \int_0^t dt' (u_x \dot{u}_y - \dot{u}_x u_y)   \\
    &  \times ( \partial u_x \bra{m_{\vec{k};\vec{u}}}\partial u_y \ket{n_{\vec{k};\vec{u}}} - \partial u_y \bra{m_{\vec{k};\vec{u}}}\partial u_x \ket{n_{\vec{k};\vec{u}}} ) .
\end{split}
\end{equation}
Similarly to how Berry curvature can be rewritten using the transformation $\bra{n}\frac{\partial H}{\partial \vec{u}}\ket{n'} 
= \braket{\frac{\partial n}{\partial \vec{u}}|n'} (\varepsilon_n - \varepsilon_{n'})$ for $n' \neq n$ \cite{Xiao2010}, we reformulate the geometric phase as
\begin{equation}\label{eq:geometric_phase_Pi}
\begin{split}
    \gamma_{mn}(t) & = \iu \int_0^t dt' (u_x \dot{u}_y - \dot{u}_x u_y)   \\
    &  \times \sum_{n'\neq m}\left[\frac{\bra{m_{\vec{k}}}\partial_{u_x}\op{H}_{\vec{u}(t)} \ket{n'_{\vec{k}}}\bra{n'_{\vec{k}}}\partial_{u_y}\op{H}_{\vec{u}(t)} \ket{n_{\vec{k}}}}{\left(\epsilon_{m\vec{k}}-\epsilon_{n'\vec{k}}\right)^2}\right.\\
    & - \left.\frac{\bra{m_{\vec{k}}}\partial_{u_y}\op{H}_{\vec{u}(t)} \ket{n'_{\vec{k}}}\bra{n'_{\vec{k}}}\partial_{u_x}\op{H}_{\vec{u}(t)} \ket{n_{\vec{k}}}}{\left(\epsilon_{m\vec{k}}-\epsilon_{n'\vec{k}}\right)^2}\right].
\end{split}
\end{equation}
By comparing with equation \eqref{eq:Pi}, we see that in the low frequency limit, i.e. $\hbar\omega \ll \abs{\epsilon_{m\vec{k}}-\epsilon_{n\vec{k}}}$, the expression inside the integral in equation is proportional to $\Pi_{xy}$ from the phonon inverse Faraday effect formalism and the phonon angular momentum, i.e. $\vec{L} = \vec{u} \times \dot{\vec{u}}$. We note that a similar expression can also be derived starting from the time-dependent Aharonov-Andan phase~\cite{aharonov1987phase} as discussed in the supplement of Ref. \cite{klebl2024ultrafast}.

\subsubsection{Current-based formalism}
Both, the magnetization $\vec{M}$ and polarization $\vec{P}$ of a sample can be summarized in a constituent equation, involving the boundary current $\vec{j}$,
\begin{equation}
    \vec{j} = \dot{\vec{P}} + \nabla\times\vec{M}.
    \label{sec:adiabatic:current}
\end{equation}
In a semiclassical approach, the boundary current can be decomposed into two contributions, $\vec{j} = \vec{j}^{(1)} +\vec{j}^{(2)}$ ~\cite{xiao2009polarization,ren2021phonon}. For the boundary current due to an ionic displacement $\tau_i$ these contributions are given by
\begin{align}
    j_i^{(1)} &= e\,\dot{\tau}_j \int_{\text{BZ}}\,\frac{\mathrm{d}\vec{k}}{\left(2\pi\right)^d}\,\Omega_{k_i \tau_j},\\
    j_i^{(2)} &= e\,\dot{\tau}_k \int_{\text{BZ}}\,\frac{\mathrm{d}\vec{k}}{\left(2\pi\right)^d}\,\Omega_{k_i k_j r_j \tau_k}.
\end{align}
where $\alpha$ denotes the Cartesian component of $\vec{j}^{(i)}$, $\delta$ denotes the Cartesian component of the displacement vector $\vec{\tau}$, $e$ is elementary charge with a positive sign, $\vec{k}$, $ \vec{r}$ are momentum and real space coordinates, respectively. Double indices are summed over. Also, we use the abbreviation $\Omega_{\alpha \beta \gamma \delta}= \Omega_{\alpha \beta}\Omega_{\gamma \delta} + \Omega_{\beta \gamma}\Omega_{\alpha \delta} - \Omega_{\alpha \gamma}\Omega_{\beta \delta}$ with the Berry curvature $\Omega_{\alpha \beta} = \partial_{\alpha}A_{\beta}-\partial_{\beta}A_{\alpha}$ and the Berry connection $A_{\alpha}=\langle \varphi|i\partial_{\alpha}|\varphi\rangle$. Note that the Berry connection is Abelian in the single occupied electron band case and becomes non-Abelian for multiple occupied bands~\cite{ren2021phonon}. For simplicity, we focus on the single band case in the following.

For a small displacement $\vec{\tau}$, one can expand the current density $\vec{j}$ in a Taylor series,
\begin{equation}
    j_i = \left.j_i\right|_{\vec{\tau}=0} + \left(\nabla_{\vec{\tau}}j_i\right)\cdot\vec{\tau} + \dots.
\end{equation}
Comparing the resulting expression with the definition of the current given in equation \eqref{sec:adiabatic:current} allows to identify the following expression for the magnetization, stemming from the topological current density $\vec{j}^{(2)}$~\cite{ren2021phonon},
\begin{equation}\label{eq:magnetization_adiabatic}
    M_z = \frac{e}{2m_{\text{I}}} L_{\text{I}} \int\frac{d\vec{k}}{(2\pi)^d}\, \Omega_{k_{x}k_{y}\tau_{x}\tau_{y}} \, ,
\end{equation}
where $m_{\text{I}}$ is the mass of the ion and $L_{\text{I}} = m_I \left(\vec{\tau}\times\dot{\vec{\tau}}\right)$ the ionic angular momentum. By definition, one can write $\Omega_{k_{x}k_{y}\tau_{x}\tau_{y}} = \Omega_{k_{x}\tau_{y}}\Omega_{k_{y}\tau_{x}}-\Omega_{k_{x}\tau_{x}}\Omega_{k_{y}\tau_{y}}+\Omega_{k_{x}k_{y}}\Omega_{\tau_{x}\tau_{y}}$, which allows to split the magnetization into two parts, $\vec{M} = \vec{M}^{(a)} + \vec{M}^{(b)}$. $\vec{M}^{(a)}$ contains the first two terms of $\Omega_{k_{x}k_{y}\tau_{x}\tau_{y}}$, while $\vec{M}^{(b)}$ is computed from the third term (we use notation $a$ and $b$ to avoid confusion with the two parts of the current given in equation \eqref{sec:adiabatic:current}). By the modern theory of the polarization, the polarization is defined in terms of the Berry connection, $P_i = \int\frac{d\vec{k}}{(2\pi)^d} A_{\vec{k}_i}$~\cite{King-smith1993}. Hence, $\vec{M}^{(a)}$ is related to the Born effective charge \eqref{eq:BornZ} per unit volume~\cite{Ghosez1998,Mele2002},
\begin{equation}
    Z_{ij} = \frac{1}{e}\frac{\mathrm{d}P_i}{\mathrm{d}\tau_j} = e \int\frac{d\vec{k}}{(2\pi)^d}\,\Omega_{k_i \tau_j},
\end{equation}
and can be written as follows
\begin{equation}
    M_z^{(a)} = \frac{e}{2m_I} L_I \int\frac{d\vec{k}}{(2\pi)^d}\, \left[\Omega_{k_{x}\tau_{y}}\Omega_{k_{y}\tau_{x}}-\Omega_{k_{x}\tau_{x}}\Omega_{k_{y}\tau_{y}}\right].
\end{equation}
The second contribution to the magnetization, $M_z^{(b)}$, is a topological contribution stemming from the boundary current of the sample,
\begin{equation}
    M_z^{(b)} = \frac{e}{2m_I} L_I \int\frac{d\vec{k}}{(2\pi)^d}\,\Omega_{k_{x}k_{y}}\Omega_{\tau_{x}\tau_{y}}
\end{equation}

Numerical studies of graphene as well as Cd$_3$As$_2$ and PbTe by Ren et al. show that this theory yields phonon magnetic moments $10^{3-5}$ times larger than the atomic magneton.

\subsection{Artificial gauge fields induced by axial phonons}\label{sec:artificial-gauge}
Strain in materials modifies the atomic distances and as a consequence the hopping amplitudes for electrons on the lattice. In Dirac materials, such as graphene, where the electronic states are effectively described by the Dirac equation~\cite{CastroNeto2009,Abergel2010,Wehling2014}, this modification induces effective gauge fields~\cite{Iorio2015,deJuan2013,Vozmediano2010,Guinea2010,Farjam2009,Pereira2009}. This concept can be generalized to phonons, dynamically straining the Dirac material. For a massless Dirac material, the resulting Hamiltonian is given by~\cite{Hu2021,Chen2025geo,Chen2025gauge,Chaudhary2025}
\begin{equation*}
    \op{H} = v_D\left(p_j - e A_j - e \chi a_j^\nu\right)\gamma^j.
\end{equation*}
Here, $v_D$ is the Dirac velocity, the matrices $\gamma^i$ are a set of mutually anticommuting Dirac matrices, $A_i$ is the electromagnetic potential, while $a_j^\mu$ are defined as
\begin{equation}
    a^\nu_i = \frac{g^\nu}{ev_D} \tau^\nu_i.
    \label{phonon_a}
\end{equation}
The parameter $g$ denotes the electron-phonon coupling strength, and $\chi = \pm 1$ is a prefactor for the two valleys in the Dirac equation. While the electromagnetic gauge field enters the Dirac Hamiltonian via the kinetic momentum $p_j \rightarrow p_j - eA_j$, the phonon modes enter in terms of a chiral gauge field.

Integrating out the fermions~\cite{Chen2025gauge,Chaudhary2025} yields an effective action for the electromagnetic and phononic gauge potentials, taking the form of a Chern-Simons theory:
\begin{align}
    S_{\text{eff}}[\vec{A}] &= \frac{\sigma_{xy}}{2} \int\mathrm{d}x\,\epsilon^{ijk} A_i \partial_j A_k, \label{S:EM} \\
    S_{\text{eff}}[\vec{a}^\nu] &= \frac{\sigma_{xy}}{2} \int\mathrm{d}x\,\epsilon^{ijk} a_i \partial_j a_k. \label{S:a}
\end{align}
For the electromagnetic potential, the effective action \eqref{S:EM} allows the Hall current to be determined by $J_i = \frac{\delta S_{\text{eff}}[\vec{A}]}{\delta A_i} = \sigma_{xy}\epsilon_{ij}E_j$. Consequently, the phonon gives rise to an alternating valley Hall current:
\begin{equation}
    J^\nu_i = \frac{\delta S_{\text{eff}}[\vec{a}^\nu]}{\delta a^\nu_i} = \sigma_{xy}\epsilon_{ij} \partial_t a^\nu_j.
\end{equation}
This term emerges for systems with broken time-reversal symmetry. Considering an optical phonon mode $\nu$ with atomic displacements $\tau_j^\nu$, substituting $a_j^\nu$ in equation \eqref{S:a} with \eqref{phonon_a} and including the kinetic and potential energy yields the phonon Lagrangian:
\begin{equation}
    \mathcal{L}^\nu = \frac{\rho_I}{2}\left[\left(\dot{\vec{\tau}}^\nu \right)^2 - \omega_\nu^2 \left(\vec{\tau}^\nu\right)^2\right] + \eta_H^{\text{ph}}\,\vec{\tau}^\nu\times\dot{\vec{\tau}}^\nu.
\end{equation}
Here, the phonon Hall viscosity is given by~\cite{Chen2025gauge}
\begin{equation}
    \eta_H^{\text{ph}} = \frac{\sigma_{xy}\left(g^{\nu}\right)^2}{2 e^2 v_D^2},
\end{equation}
with $\rho_I$ being the ionic density. The phonon angular momentum term ($\sim \vec{\tau}^\nu \times \dot{\vec{\tau}}^\nu$) gives rise to a Lorentz force on the phonons, arising from the interaction with the electrons:
\begin{equation}
    \vec{F} = \frac{\delta \mathcal{L}^\nu}{\delta \vec{\tau}^\nu} = \eta_H^{\text{ph}}\,\hat{\vec{z}}\times \dot{\vec{\tau}}^\nu.
\end{equation}
Equations of motion for left and right circularly polarized phonon modes $\tau^\nu_\pm = \frac{1}{\sqrt{2}}\left[\tau_x^\nu \pm \iu \tau_y^\nu \right]$ are:
\begin{equation}
    \rho_I\left(\ddot\tau^\nu_{\pm} + \omega_\nu^2 \tau^\nu_ \pm\right) = \pm 2\iu \eta_H^{\text{ph}} \tau_\pm^\nu.
\end{equation}
Thus, the Hall viscosity induces a splitting of the frequencies for phonon modes $\omega^\nu_\pm$:
\begin{equation}
    \omega_\pm^\nu = \sqrt{\omega_\nu^2 + \omega_\eta^2} \pm \omega_\eta,
    \label{Hall:frequency}
\end{equation}
with $\omega_\eta = \eta_H^{\text{ph}}/\rho_I$. Connecting equation \eqref{Hall:frequency} with the definition of the phonon magnetic moment yields the following form~\cite{Chen2025gauge,Chen2025geo,Chaudhary2025}:
\begin{equation}
    \mu_{\text{ph}} = \frac{\left(g^{\nu}\right)^2}{v_D^2 \rho_I B} \frac{\hbar}{2 e^2} \sigma_{xy}.
\end{equation}
This approach reproduces, for example, the experimentally observed phonon Zeeman splitting in the Dirac material Cd$_3$As$_2$~\cite{cheng2020large,Chen2025gauge}.

\subsection{Floquet approach}\label{subsec:floquet_Klebl}
Floquet theory applies to systems with a time-periodic potential, and the Hamiltonian satisfying $\op{H}(t + T) = \op{H}(t)$. In this case the solutions of the Schrödinger equation can be written using the Floquet theorem $\ket{\psi(t)} = e^{-\iu \frac{\epsilon_\alpha}{\hbar} t}\ket{\phi_\alpha(t)}$~\cite{Rodriguez-Vega2018,Oka2019}, giving rise to the Floquet Schrödinger equation
\begin{equation}
    \left[\op{H}(t) - \iu \hbar \frac{\partial}{\partial t}\right] \ket{\phi_\alpha(t)} = \epsilon_\alpha \ket{\phi_\alpha(t)}.
\end{equation}
Here $\epsilon_\alpha$ takes the role of a quasienergy and $\ket{\phi_\alpha(t)}$ is a $T$-periodic function $\ket{\phi_\alpha(t+T)} = \ket{\phi_\alpha(t)}$. As a consequence, $\ket{\phi_\alpha(t)}$ can be expressed in terms of a Fourier series in the frequency $\omega = 2\pi/T$,
\begin{equation}
    \ket{\phi_\alpha(t)} = \sum_n e^{\iu n \omega t} \ket{\phi_{\alpha; n}}.
    \label{floquet:fourrier}
\end{equation}
An undamped axial phonon mediates the time-periodic potential. For example, we assume a phonon mode
\begin{equation}
    \vec{u}= u (\cos{(\omega t),\sin{(\omega t)}},0)^{T}\,,    
\end{equation}
together with a two level system, coupled by the Hamiltonian
\begin{equation}
    \op{H}_c = g\,\vec{u}(t)\cdot\vec{\sigma} = g u \left(\sigma_+ e^{-\iu\omega t} + e^{\iu t}\sigma_-\right).
    \label{Floquet:Ham}
\end{equation}
The Hamiltonian \eqref{Floquet:Ham} has been motivated, e.g., in Ref.~\cite{klebl2024ultrafast}, representing the electronic subspace spanned by $p$-orbitals $\ket{p_+}$ and $\ket{p_-}$ in the oxide perovskite SrTiO$_3$, with an electron-phonon coupling mediated by local Jahn-Teller distortions. The electron phonon coupling strength is given by $g$. Taking matrix elements over $\op{H}_c$ involving the Fourier expansion \eqref{floquet:fourrier} gives the following form of the Floquet Schrödinger equation~\cite{Vogl2020}
 \begin{multline}
\begin{pmatrix}
\ddots & \vdots & \vdots & \vdots & \vdots & \vdots & & \\
\cdots & \op{h}^{\dagger} & \op{h}_0 - \hbar\omega & \op{h} & 0 & 0 & \cdots \\
\cdots & 0 & \op{h}^{\dagger} & \op{h}_0 & \op{h} & 0 & \cdots \\
\cdots & 0 & 0 & \op{h}^{\dagger} & \op{h}_0 + \hbar\omega & \op{h} & \cdots \\
& \vdots & \vdots & \vdots & \vdots & \vdots & \ddots
\end{pmatrix}
\begin{pmatrix}
\vdots \\
 \ket{\phi_{\alpha; -1}} \\
 \ket{\phi_{\alpha; 0}} \\
 \ket{\phi_{\alpha; 1}} \\
\vdots
\end{pmatrix}
\\= \epsilon_\alpha \begin{pmatrix}
\vdots \\
\ket{\phi_{\alpha; -1}} \\
 \ket{\phi_{\alpha; 0}} \\
 \ket{\phi_{\alpha; 1}} \\
 \vdots
\end{pmatrix}.
\end{multline}
Here, we defined $\op{h} = \frac{1}{T} \int \mathrm{d}t\, e^{-\iu \omega t} \op{H}_c = g u \sigma_-$ and $\op{h}^\dagger = g u \sigma_+$. Resolving for each individual component $\ket{\phi_{\alpha;n}}$ gives rise to a continued fraction~\cite{Perfetto2015,Giovannini2020}. For weak driving, this fraction can be truncated after first order, giving rise to an effective Floquet Hamiltonian~\cite{Vogl2020,Perfetto2015,Giovannini2020}
\begin{equation}
    \label{eq:Klebl-eff-interaction}
    \op{H}^{\text{eff}} = \op{h}_0 + \op{h} \frac{1}{\epsilon_\alpha - \op{h}_0 - \hbar\omega} \op{h}^{\dagger} + \op{h}^{\dagger} \frac{1}{\epsilon_\alpha - \op{h}_0 + \hbar\omega} \op{h} \, .
\end{equation}
In the high-frequency limit, $\omega \gg \epsilon_\alpha$, the quasienergy can be disregarded~\cite{klebl2024ultrafast} and one ends up with a conventional Schrödinger equation. Since the Pauli matrices act on the space spanned by $\ket{p_+}$ and $\ket{p_-}$, terms proportional to $\sigma_z$ in $\op{H}^{\text{eff}}$ lead to an energy splitting of the $p_+$ and $p_-$ state. This allows to identify an effective magnetic field, given by~\cite{klebl2024ultrafast}
\begin{equation}
    B^{\text{eff}} = \frac{1}{\mu_B} \frac{\operatorname{Tr}\left[\sigma_z \op{H}^{\text{eff}}\right]}{2} = - \frac{g^2 u^2}{\mu_B \hbar\omega}.
\end{equation}
In contrast to the perturbation theory and the adiabatic approach, this equation does not involve the band gap in the denominator.

\subsection{Inertial effects}\label{subsec:intertial}
Inertial effects emerge in accelerating frames of reference. While inertial effects have been intensively studied in both classical~\cite{misner1973gravitation} and quantum mechanics~\cite{strange2016dirac,ryder2008spin,hehl1985kinematics,hehl1990inertial,muller1976two}, they were only recently discussed as a potential mechanism for coupling electronic spin with rotations in molecules and crystals~\cite{geilhufe2022inertial,Matsuo2013,Matsuo2011b,Matsuo2011}.

\begin{figure*}
    \centering
    \includegraphics[width=0.95\textwidth]{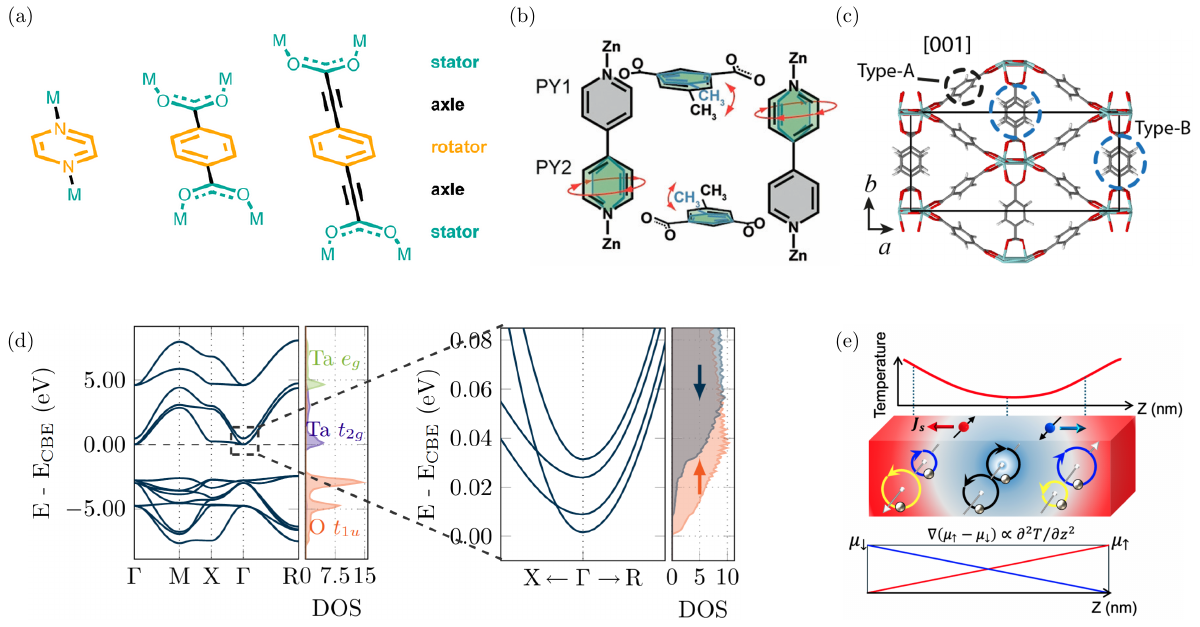}
    \caption{Spin-rotation coupling (microscopic Barnett effect) in systems with rotational degrees of freedom. (a) Schematics of molecules as platforms to host rotators. (b) Rotational degrees of freedom in the metal-organic framework Zn(5-Me-isophthalate)(bipyridine). (c) Rotational degrees of freedom in the metal-organic framework MIL-140A. (d) Expected spin splitting of energy bands due to spin-rotation coupling induced by axial phonons. (e) Spin-rotation coupling as the microscopic mechanism to explain spin transport in a hybrid organic-inorganic perovskite with a parabolic temperature profile. \textit{(a, reproduced from~\cite{Gonzalez-Nelson2019} under the Creative Commons Attribution license)} \textit{(b, reproduced with permission from~\cite{Inukai2018}; Copyright 2018 Wiley-VCH.)} \textit{(c, reproduced with permission from~\cite{ryder2017}; Copyright 2017 American Physical Society.)} \textit{(d, reproduced from~\cite{geilhufe2023KTO} under the Creative Commons Attribution license)} \textit{(e, reproduced with permission from~\cite{Qin2025}; Copyright 2025 American Physical Society.)}}
    \label{fig:rotations}
\end{figure*}

Classically, this coupling can be motivated by the Coriolis force acting on a probe particle (electron) in a rotating reference frame, $\vec{F}_{\text{Coriolis}} = 2\vec{p} \times \vec{\omega}$. Here, $\vec{\omega}$ is the angular velocity of the reference frame (e.g., circular motion of ions in chiral phonons), and $\vec{p}$ is the momentum of the probe particle. Promoting the force to an energy by multiplying with the position $\vec{r}$ yields $E = \vec{r} \cdot \vec{F}_{\text{Coriolis}} = \vec{\omega} \cdot \vec{L}$, where $\vec{L}$ is the electron angular momentum. The extension to quantum mechanics gives rise to the spin-rotation term, or Barnett field, in the Hamiltonian,
\begin{equation}
    \op{H} = \vec{\omega} \cdot \op{\vec{J}},
    \label{eq:spin-rotation}
\end{equation}
where $\op{\vec{J}} = \op{\vec{L}} + \op{\vec{S}}$ is the total angular momentum of the electron, combining orbital angular momentum $\op{\vec{L}}$ and spin $\op{\vec{S}}$.

More rigorously, inertial effects are described quantum mechanically by formulating the Dirac equation in the accelerating frame,
\begin{equation}
    \gamma^a \mathrm{i}\hbar D_a \Psi = mc \Psi,
    \label{codev}
\end{equation}
where $\Psi$ is a four-spinor, $m$ is the electron mass, and $\gamma^a$ are the Dirac $\gamma$-matrices, satisfying $\{\gamma^\mu, \gamma^\nu\} = \eta^{\mu\nu}$. The covariant derivative is given by $D_a = \partial_a - \frac{\mathrm{i}}{4} \sigma^{bc} \Gamma_{bca}$, with $\sigma^{ab} = \frac{\mathrm{i}}{2}[\gamma^a, \gamma^b]$ and $\Gamma_{bca}$ the connection coefficient~\cite{hehl1985kinematics,hehl1990inertial}. For circular motion of ions, the Dirac equation~\eqref{codev} can be rewritten as~\cite{hehl1990inertial,geilhufe2022inertial}
\begin{multline}
    \mathrm{i}\hbar \partial_t \Psi = \left[c \vec{\alpha} \cdot \vec{p} - \frac{\gamma^2}{2mc} \left\{\vec{F}_{\text{Cent.}} \cdot \vec{r}, \vec{p} \cdot \vec{\alpha}\right\} \right. \\
    \left. + \beta \left(mc^2 - \gamma^2 \left(\vec{F}_{\text{Cent.}} \cdot \vec{r}\right)\right) - \vec{\omega} \cdot \vec{J}\right] \Psi,
    \label{deq}
\end{multline}
where $\gamma = \left(1 - \frac{\vec{\tau}^2 \omega^2}{c^2}\right)^{-\frac{1}{2}} \approx 1$ is the Lorentz factor, and $\vec{F}_{\text{Cent.}} = m\omega^2 \vec{\tau}$ is the centrifugal force, with $\vec{\tau}$ being the ionic displacement. Equation~\eqref{deq} can be reduced to the non-relativistic limit using the Foldy-Wouthuysen transformation~\cite{bjorken1964relativistic}, yielding four terms summarized in Table~\ref{inertial}~\cite{geilhufe2022inertial}.

Besides the spin-rotation coupling corresponding to the Coriolis force, three additional terms emerge due to uniaxial symmetry breaking induced by the centrifugal force: a centrifugal field coupling $\sim \gamma^2 \vec{F}_{\text{Cent.}} \cdot \vec{r}$, a redshift term $\sim \vec{p} \left(\vec{F}_{\text{Cent.}} \cdot \vec{r}\right) \vec{p}$, and an inertial Rashba-type spin-orbit interaction $\sim \vec{F}_{\text{Cent.}} \cdot \left(\vec{S} \times \vec{p}\right)$. While the latter two are strongly suppressed by $\sim (mc)^{-2}$, the centrifugal field coupling is generally weak, comparable to the coupling of electrons to an electric field. In this analogy, the effective electric field mediated by the centrifugal force is on the order of $10~\text{V cm}^{-1}$. In contrast, the spin-rotation coupling corresponding to the Coriolis force can introduce a splitting of spin-degenerate bands on the order of $10~\text{meV}$ for typical frequencies of chiral phonons~\cite{geilhufe2023KTO}, and this splitting increases under strong spin-orbit interaction.

\begin{table}[]
    \centering
    \begin{tabular}{ll}
        \hline\hline
        Spin-rotation coupling / Barnett field & $\vec{\omega} \cdot \vec{J}$ \\
        Centrifugal field coupling & $\gamma^2 \vec{F}_{\text{Cent.}} \cdot \vec{r}$ \\
        Centrifugal spin-orbit coupling & $\frac{\gamma^2}{2m^2c^2} \vec{F}_{\text{Cent.}} \cdot \left(\vec{S} \times \vec{p}\right)$ \\
        Centrifugal redshift & $\frac{\gamma^2}{2m^2c^2} \vec{p} \left(\vec{F}_{\text{Cent.}} \cdot \vec{r}\right) \vec{p}$ \\
        \hline\hline
    \end{tabular}
    \caption{Inertial effects of quantum systems in rotating reference frames~\cite{geilhufe2022inertial}.}
    \label{inertial}
\end{table}

Inertial effects are universal in accelerated frames of reference. The influence of inertial effects on the quantum mechanical wave function has been verified using neutron interferometry~\cite{Page1975,Werner1979,mashhoon1988neutron}, detecting interference caused by Earth's rotation ($1/\text{day} \approx 11~\mu\text{Hz}$). Similarly, the spin-rotation coupling was also observed when one of the neutron beams passes through a rotating magnetic field with frequencies in the kHz range~\cite{Danner2020}.

The existence of inertial effects becomes particularly promising in the nanoscale regime, where rotation frequencies are much higher. One example is the application of nano-resonators to replace magnetic fields in spintronic devices~\cite{Matsuo2011,Matsuo2011b,Matsuo2013}. Rotational degrees of freedom also naturally arise in metal-organic framework materials (MOFs), as shown in \autoref{fig:rotations}(a)--(c). MOFs are organic-inorganic hybrid materials~\cite{yaghi1995selective}, composed of inorganic building units linked by organic molecules~\cite{Eddaoudi2002}. Their structures are typically sparse, allowing for freely rotating molecules~\cite{Gonzalez-Nelson2019} at sufficiently high temperatures (often room temperature). Rotational modes in MOFs can also be coherently excited using THz laser light~\cite{ryder2017} (\autoref{fig:rotations}(c)). These molecular rotations are conjectured to couple to the electronic angular momentum via the spin-rotation coupling (microscopic Barnett effect), as described by equation~\eqref{eq:spin-rotation}. Axial excitations in MOFs and the resulting rotomagnetic response have been discussed as a promising path toward quantum sensing, e.g., in connection with dark matter detection~\cite{romao2023chiral}. Beyond rotations, MOFs host other unconventional degrees of freedom (e.g., buckling, porosity, interpenetration, framework topology), making these materials promising quantum materials~\cite{Huang2024}.

On an even smaller scale, inertial effects—and in particular, the spin-rotation coupling—have also been discussed for chiral and axial phonons, where the ionic motion is treated semiclassically~\cite{geilhufe2022inertial,geilhufe2023KTO,Qin2025}. For example, in KTaO$_3$, the coherent and circularly polarized excitation of the ferroelectric soft mode ($\omega \approx 2.4~\text{THz}$ at 300~K) introduces a spin splitting in the spin-orbit split conduction bands on the order of $\approx 10~\text{meV}$ (\autoref{fig:rotations}(d)). Probing the spin splitting of electronic states using spectroscopic methods would thus provide strong evidence for the existence of this effect. Furthermore, the spin-rotation coupling, or microscopic Barnett effect, has been used to compute spin-transport properties of hybrid organic-inorganic perovskites~\cite{Qin2025} in a parabolic temperature profile (\autoref{fig:rotations}(e)).

\section{Discussion and outlook}

We have presented an overview of the current status on investigating the interplay of axial phonons and magnetism. Various microscopic theories were brought forward motivating the coupling of phonon angular momentum and magnetic degrees of freedom. While many such theories were developed on specific examples, we can see the formal agreement in the final expressions for the phonon magnetic moment, independently of the chosen method, such as perturbation theory, Floquet theory, adiabatic motion, inertia or pseudo-gauge fields. Such a universal agreement on the shape of the final expression is not surprising from a symmetry perspective, dictating the formation of a scalar quantity from time-reversal symmetry breaking vectors. Combined with the experimental evidence for significant magnetization as shown in section \ref{section:experiments}, axial phonons promise a new way to manipulate the magnetic properties of materials. Their potential use for controlling magnetic order has already been observed experimentally \cite{Davies2024}, motivating application areas involving data storage, data processing and spintronics. As the field of axial phono-magnetism is evolving rapidly, we see challenges and open questions ahead.

First, despite promising experimental signatures, the physical nature of the effective magnetic field induced by axial phonons is still under debate. It has been proposed that the field induced by circularly polarized phonons is not a magnetic field obeying Maxwell's equations, but only mimics its effects inside the material \cite{merlin2023unraveling, merlin2025magnetophononics}. Indeed, the experiments discussed in section \ref{section:experiments} do not measure the magnetic field directly, but through other variables, e.g. the size of the phonon Zeeman splitting or probe polarization rotation in the magneto-optical Kerr effect. 
It is reasonable to suggest that these effects are a consequence of the time reversal symmetry breaking, and not the magnetic field itself.
However, the magnetic switching experiment \cite{Davies2024} suggests that axial phonons have the capacity to influence magnetic properties even outside of the sample, which seemingly contradicts the idea that the phonon-induced field is undetectable outside of the sample. Hence, it would be required to conduct additional experiments, distinguishing between a pesodomagnetic field and a physical Maxwellian magnetic field, e.g., by placing a magnetic sensor in the vicinity of the sample, without a direct interface. 

Another open question relevant for applications of phono-magnetic effects is which materials or groups of materials are the best candidates for maximizing the size of the effect. Various theoretical approaches discussed in this review derived an effective magnetic field depending on $\Delta^{-2}$, with $\Delta$ being the size of the gap. Therefore, materials with a small gap would be a promising choice for maximizing the size of the phono-magnetic effects. However, it is important to stress that experiments on wide gap materials still give a considerable effect~\cite{Basini2024,Davies2024}. Additionally, as discussed in section \ref{subsec:exp_phonon_Zeeman}, experimental evidence suggests that the phonon magnetic moment increases in topologically nontrivial materials. Thus, studying axial phonons in topological insulators would be a promising route to understand the relationship between topology and phonon-induced magnetism. Additionally, an extension to soft materials, organics and framework materials can lead to elevated phonon angular momenta or pure rotational degrees of freedom.

Third, due to time-reversal symmetry breaking, magnetic materials naturally host axial phonons. However, these phonons also play a role in enhancing the magnetism~\cite{fransson2025chiral}. Linking the magnetic phase diagram, determined by Néel and Curie temperatures, with the emergence of axial phonons and the size of the phonon magnetic moment would be another important puzzle piece. This direction is also closely conneted to investigating the coupling of axial phonons and magnons~\cite{weissenhofer2024truly, fransson2025chiral, Miranda2025, bonini2023frequency}.

A fourth possible direction would be to investigate whether other charged quasiparticles, such as polaritons~\cite{Yaniv2025}, can possess angular momentum and induce magnetic effects similar to axial phonons. An extension of this direction would be the development of a more general theory that would describe the magnetic effects from charged quasiparticles. Evidence for this direction is given by discussing metamaterials, such as skyrmion lattices, where similar coupling mechanisms between Berry connection and angular momentum have been brought forward~\cite{Benzoni2021,Marijanovi2022}. 
        
A fifth direction comprises of the role of the axial phonomagnetism in transport experiments, i.e., the phonon thermal Hall effect. Recent experimental observations demonstrate a large thermal Hall conductivity that is theorized to originate from the chirality of phonons \cite{grissonnanche2019giant, grissonnanche2020chiral}. At the same time, transport of chiral phonons is shown to generate Hall viscosity, which is in turn proportional to the external TRS-breaking magnetic field \cite{Flebus2023Phonon,Behnia2025}. Considering the distinction between chiral and axial phonons, it is promising to investigate the relationship of the magnetic fields induced by axial achiral phonons and a non-zero Hall viscosity. 

Thus, we can conclude that significant progress on phono-magnetic effects facilitates a multitude of exciting and promising research directions in the upcoming years. 

\begin{acknowledgments}
We are grateful for discussions with Alexander V. Balatsky, Martina Basini, Stefano Bonnetti, Swati Chaudhary, Michael Fechner, Dominik Juraschek, Ylva Liljegren, and Hanyu Zhu. 
We acknowledge support from the Swedish Research Council (VR starting Grant No. 2022-03350), the Olle Engkvist Foundation (Grant No. 229-0443), the Royal Physiographic Society in Lund (Horisont), the Knut and Alice Wallenberg Foundation (Grant No. 2023.0087), and Chalmers University of Technology, via the department of physics and the Areas of Advance Nano and Materials Science. 
\end{acknowledgments}

\bibliography{references}
\end{document}